\documentclass{article}

\usepackage{arxiv}

\usepackage[utf8]{inputenc} 
\usepackage[T1]{fontenc}    
\usepackage{hyperref}       
\usepackage{url}            
\usepackage{booktabs}       
\usepackage{amsfonts}       
\usepackage{nicefrac}       
\usepackage{microtype}      
\usepackage{cleveref}       
\usepackage{graphicx}
\usepackage[numbers,square]{natbib}
\usepackage{doi}
\usepackage{multirow}
\usepackage[table,xcdraw]{xcolor}
\usepackage{float}
\usepackage{longtable}

\title{The Architecture Tradeoff and \\ Risk Analysis Framework (ATRAF):\\\large A Unified Approach for Evaluating Software Architectures,\\ Reference Architectures, and Architectural Frameworks}

\date{}

\usepackage{authblk}

\newbox{\orcid}\sbox{\orcid}{\includegraphics[scale=0.06]{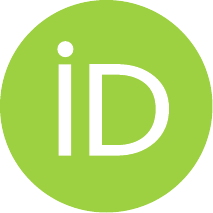}}

\author[1,2,3]{%
	\href{https://orcid.org/0009-0005-8915-7905}{\usebox{\orcid}\hspace{1mm}Amine Ben~Hassouna\thanks{\texttt{amine.benhassouna@medtech.tn, amine.benhassouna@dracodes.com (Corresponding author)}}}%
}

\affil[1]{Mediterranean Institute of Technology, South Mediterranean University, Tunis, Tunisia}
\affil[2]{Dracodes, Tunis, Tunisia}
\affil[3]{ENSI - École Nationale des Sciences de l’Informatique, Manouba, Tunisia}

\renewcommand{\headeright}{Preprint}
\renewcommand{\undertitle}{Preprint}
\renewcommand{\shorttitle}{Architecture Tradeoff and Risk Analysis Framework (ATRAF)}

\hypersetup{
  pdftitle={Architecture Tradeoff and Risk Analysis Framework (ATRAF)},
  pdfsubject={cs.SE},
  pdfauthor={Amine Ben Hassouna},
  pdfkeywords={Software architecture, Architectural frameworks, Tradeoff analysis, Quality attributes, Architecture evaluation},
}

\begin{document}
\maketitle

\begin{abstract}
Modern software systems are guided by hierarchical architectural concepts—software architectures, reference architectures, and architectural frameworks—each operating at a distinct level of abstraction. These artifacts promote reuse, scalability, and consistency, but also embed tradeoffs that shape critical quality attributes such as modifiability, performance, and security. Existing evaluation methods, such as the Architecture Tradeoff Analysis Method (ATAM), focus on system-specific architectures and are not designed to address the broader generality and variability of higher-level architectural forms. To close this gap, we introduce the Architecture Tradeoff and Risk Analysis Framework (ATRAF)—a unified, scenario-driven framework for evaluating tradeoffs and risks across architectural levels. ATRAF encompasses three methods: the Architecture Tradeoff and Risk Analysis Method (ATRAM), extending ATAM with enhanced risk identification for concrete systems; the Reference Architecture Tradeoff and Risk Analysis Method (RATRAM), adapting ATRAM to the evaluation of domain-level reference architectures; and the Architectural Framework Tradeoff and Risk Analysis Method (AFTRAM), supporting the evaluation of architectural frameworks that guide entire system families. All three methods follow an iterative spiral process that enables the identification of sensitivities, tradeoffs, and risks while supporting continuous refinement of architectural artifacts. We demonstrate ATRAF through progressively abstracted examples derived from the Remote Temperature Sensor (RTS) case, originally introduced in the ATAM literature. ATRAF equips architects, reference modelers, and framework designers with a practical, systematic approach for analyzing design alternatives and managing quality attribute tradeoffs early in the lifecycle and across all levels of architectural abstraction.
\end{abstract}



\section{Introduction}
\label{sec:introduction}

Modern software systems increasingly rely on structured architectural concepts to manage complexity, enable reuse, and guide quality-driven design. These concepts---software architectures, reference architectures, and architectural frameworks---span a hierarchy of abstraction levels. Each plays a distinct role in shaping design decisions and quality attribute outcomes across the system lifecycle. At higher abstraction levels, reference architectures and frameworks aim to standardize and accelerate system development through reusable patterns, processes, and constraints. However, these architectural forms also embed tradeoffs that impact modifiability, performance, scalability, and other critical concerns across families of systems.

While methods such as the Architecture Tradeoff Analysis Method (ATAM) have significantly improved our ability to evaluate concrete software architectures, traditional approaches have not kept pace with the increasing abstraction and generality of architectural artifacts. Reference architectures and, more acutely, architectural frameworks introduce domain variability, stakeholder diversity, and long-lived applicability across multiple contexts. These characteristics present challenges for scenario elicitation, tradeoff identification, and risk analysis---challenges that conventional evaluation techniques were not designed to address.

To fill this methodological gap, we propose the \textit{Architecture Tradeoff and Risk Analysis Framework (ATRAF)}---a unified, multi-level framework for evaluating architectural decisions and quality attribute tradeoffs across the full abstraction hierarchy. ATRAF offers a comprehensive solution by integrating flexibility management, meta-level quality attributes, and multi-context scenario realization. It comprises three integrated methods: the \textit{Architecture Tradeoff and Risk Analysis Method (ATRAM)}, which extends ATAM with enhanced support for risk identification in concrete systems; the \textit{Reference Architecture Tradeoff and Risk Analysis Method (RATRAM)}, which adapts ATRAM for domain-level reuse and variation; and the \textit{Architectural Framework Tradeoff and Risk Analysis Method (AFTRAM)}, designed to evaluate the flexibility, constraints, and lifecycle support embedded in architectural frameworks. Each method follows a scenario-driven, spiral evaluation process that supports the iterative refinement of architectural artifacts while enabling the systematic identification of tradeoffs, sensitivity points, and risks across stakeholder concerns and quality attributes.

This paper makes the following 6 contributions:
\begin{itemize}
    \item \textbf{A conceptual differentiation} between software architectures, reference architectures, and architectural frameworks, highlighting their respective roles in shaping quality attribute tradeoffs.
    \item \textbf{The design and rationale} for the ATRAF framework, enabling structured tradeoff and risk analysis across architectural abstraction levels.
    \item \textbf{ATRAM}, a method for evaluating concrete software architectures, focusing on system-level tradeoffs and risks.
    \item \textbf{RATRAM}, a method for evaluating reference architectures, addressing domain-specific variability and reuse.
    \item \textbf{AFTRAM}, a method for evaluating architectural frameworks, focusing on extensibility, adaptability, and lifecycle management across system families.
    \item \textbf{A cohesive case example family} (RTSA, RTSRA, RMAF) derived from the Remote Temperature Sensor (RTS) system, demonstrating the application of ATRAF at each level of abstraction.
\end{itemize}

The remainder of this paper is structured as follows: Section~\ref{sec:hierarchical-structuring-and-differentiation-of-architectural-concepts} presents the hierarchical differentiation of architectural concepts. Section~\ref{sec:evaluating-abstract-architectures} outlines the limitations of existing methods and the need for multi-level evaluation. Section~\ref{sec:methodology} describes the methodology used to construct ATRAF. Sections~\ref{sec:atram}, \ref{sec:ratram} and \ref{sec:aftram} introduce ATRAM, RATRAM, and AFTRAM respectively, each with their evaluation phases and illustrative examples. Section~\ref{sec:cross-method-synthesis-and-lessons-learned} provides a cross-method synthesis and insights from the case family. Section~\ref{sec:related-work} reviews related work in architecture evaluation. Section~\ref{sec:conclusion} concludes the paper and discusses future directions. A detailed description of the example architectures is included in the Appendix~\ref{app:example-of-architectural-artifacts}.


\section{Hierarchical Structuring and Differentiation of Architectural Concepts}
\label{sec:hierarchical-structuring-and-differentiation-of-architectural-concepts}

\begin{figure}[ht]
    \centering
    \includegraphics[width=0.8\textwidth]{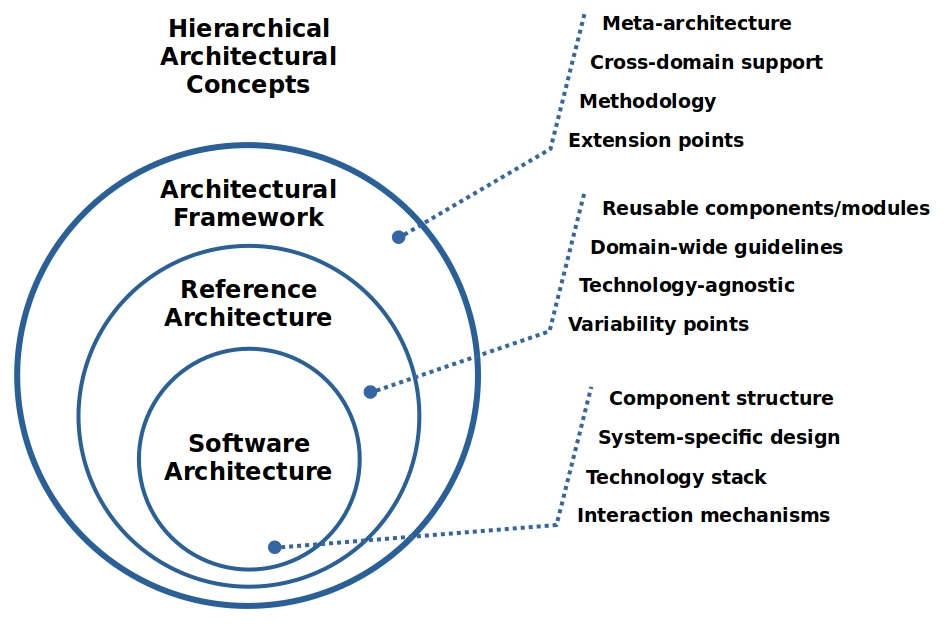}
    \caption{Hierarchical Architectural Concepts}
    \label{fig:hierarchical-architectural-concepts}
\end{figure}

Recognizing and systematically organizing the hierarchical relationships among \textbf{Software Architecture}, \textbf{Reference Architecture}, and \textbf{Architectural Framework} is critical in system design. Each represents a distinct architectural concept positioned at a specific level of abstraction and purpose. In this section, we apply the \textbf{Goals-Inputs-Outcomes (GIO)} model to structure and differentiate these concepts, facilitating a rigorous comparison of their respective goals, inputs, and outcomes. This methodological approach clarifies the key features, distinguishing characteristics, and specifications of each concept. It highlights the progression from concrete to increasingly abstract architectural forms, as depicted in Figure~\ref{fig:hierarchical-architectural-concepts}.

\subsection{Software Architecture}
\label{sec:software-architecture}

\textbf{Software Architecture} represents the most concrete level of system design. It focuses on the architecture of a specific software system, including decisions on components, technologies, and their interactions. Software architecture translates high-level principles into a functional, system-specific design that satisfies both functional and quality requirements.

\newpage

Key features of Software Architecture:
\begin{itemize}
    \item \textbf{System-specific design}: Focuses on the architecture of a particular system tailored to its specific functional and non-functional requirements.

    \item Describes \textbf{specific components} (e.g., databases, services, user interfaces) and their \textbf{interactions} (e.g., data flows, APIs).

    \item \textbf{Technological details}: Specifies technologies (e.g., SQL database, REST API) and the design of components within the system.

    \item \textbf{Custom solutions}: Optimized to meet the specific needs of the system, such as performance, security, and availability.
\end{itemize}

We illustrate this concept with the Remote Temperature Sensor Architecture (RTSA), originally developed to demonstrate the ATAM methodology. RTSA specifies the design of a system for remotely monitoring furnace temperatures, fulfilling both functional and quality requirements such as performance, availability, and security. It organizes responsibilities among three main entities---furnaces, a central server, and operator clients---with the server managing data acquisition, dynamic update scheduling, and client communication through dedicated modules and tasks. RTSA incorporates quality-driven mechanisms such as dynamic update scheduling and bounded latency communication: each furnace’s temperature is periodically read, processed through an analog-to-digital conversion module, and forwarded to clients with predictable timing. It concretely instantiates critical quality attributes through its modular structure, scheduling strategies, and basic protections against message tampering. (For a detailed description of RTSA, see Appendix~\ref{app:rtsa-description})

\subsection{Reference Architecture}
\label{sec:reference-architecture}

\textbf{Reference Architecture} operates at a broader level of abstraction than software architecture. It provides practical design guidance by generalizing architectural solutions across a domain rather than focusing on a single system. Reference architectures capture reusable patterns and domain expertise, serving as a foundational template from which multiple system-specific architectures can be derived.

\newpage

Key features of Reference Architecture:
\begin{itemize}
    \item Contains \textbf{reusable patterns} and \textbf{best practices} that can be adapted across multiple systems within a domain.

    \item Focuses on providing \textbf{domain-specific guidelines} that consolidate experience and knowledge.

    \item Offers \textbf{component models} and \textbf{architectural patterns} tailored to fit particular system requirements and evolving technologies.

    \item \textbf{Technology-agnostic}: Provides a high-level blueprint that can be realized using different platforms and technologies.

    \item \textbf{Supports evolvability}: Guides the design of future systems by consolidating proven domain-specific principles into a reusable architectural foundation.
\end{itemize}

The Remote Temperature System Reference Architecture (RTSRA)\footnote{The Remote Temperature System Architecture (RTSA), introduced in Section~\ref{sec:software-architecture}, can be instantiated from the broader RTSRA.} exemplifies a reference architecture designed for remote temperature monitoring across various industries. RTSRA provides adaptable patterns and guidelines, supporting diverse sensor types such as furnaces, air conditioning systems, and environmental sensing. It organizes core system components---temperature sources, servers, and clients---along with protocols for data acquisition, fault tolerance, and communication. RTSRA emphasizes flexibility and scalability, allowing for non-redundant or redundant server configurations and customizable fault recovery strategies. This reference architecture balances structure and flexibility, promoting reuse while enabling adaptation to specific operational needs and quality attribute goals such as availability, modifiability, and security. (For a detailed description of RTSRA, see Appendix~\ref{app:rtsra-description})

\subsection{Architectural Framework}
\label{sec:architectural-framework}

\textbf{Architectural Framework} stands at an even higher level of abstraction than reference architectures. Rather than prescribing specific structures, it provides a structured methodology and flexible principles for developing architectures across diverse domains and system families. Frameworks can encompass multiple reference architectures and extend beyond individual domains to support evolving needs.

Key features of Architectural Framework:
\begin{itemize}
    \item \textbf{Process-driven}: Defines the overall approach for designing architectures, including methodologies, decision-making processes, and lifecycle considerations; organizes work into structured \textbf{viewpoints} to address stakeholder concerns.

    \item \textbf{Encompasses reference architectures}: Provides flexibility to incorporate reusable, domain-specific architectural models.

    \item \textbf{Flexible and adaptable}: Accommodates diverse patterns, models, and technologies.

    \item \textbf{Complete lifecycle view}: Addresses system design, maintenance, evolution, and continuous improvement.

    \item \textbf{Supports tradeoff analysis}: Enables systematic evaluation of alternative design choices and their impact on critical quality attributes.
\end{itemize}

For illustration, we crafted the Remote Monitoring Architectural Framework (RMAF)\footnote{The Remote Temperature System Reference Architecture (RTSRA), introduced in Section~\ref{sec:reference-architecture}, can be viewed as an instance governed by the broader RMAF. While RTSRA focuses on temperature monitoring systems, RMAF provides a general methodology applicable to a wide range of monitoring solutions, including system performance, health metrics, and environmental sensing.}, a flexible, comprehensive methodology designed to support the development of remote monitoring systems across diverse domains. RMAF defines reusable building blocks, interaction models, and extensibility mechanisms for creating both domain-specific reference architectures and concrete system architectures. The framework guides architects through critical decisions on system design, tradeoff analysis, and quality attribute modeling, while addressing varying needs for flexibility, scalability, and adaptability in applications ranging from industrial monitoring to healthcare and environmental sensing. RMAF enables the structured evolution of monitoring systems, from basic sensor data collection to complex, fault-tolerant architectures. (For a detailed description of RMAF, see Appendix~\ref{app:rmaf-description}.)

\subsection{Comparative Overview}
\label{sec:architectural-abstraction-comparative-overview}

In this section, we compare \textbf{Software Architecture}, \textbf{Reference Architecture}, and \textbf{Architectural Framework} through the \textbf{Goals-Inputs-Outcomes (GIO)} model.

Table~\ref{tab:gio-comparison} presents a comparative analysis of the goals, inputs, and outcomes for each concept:

\begin{table}[ht]
    \centering
    \begin{tabular}{|p{3.5cm}|p{3.5cm}|p{3.5cm}|p{3.5cm}|}
    \hline
    \textbf{Concept} & \textbf{Goals} & \textbf{Inputs} & \textbf{Outcomes} \\ \hline
    Software Architecture & Optimize a system for specific functional and quality requirements & Concrete business requirements, technology constraints, quality requirements, and current system limitations. & Detailed architectural design, component structure, interaction mechanisms, technology stack, and rationale for key design choices.\\ \hline
    Reference Architecture & Promote reuse and consistency across multiple systems within a domain & Domain knowledge, design problems, best practices, and interoperability requirements & Generalized architectural template, modular components, reusable patterns, and guidelines for design decisions. \\ \hline
    Architectural Framework & Establish a common structure and process for architecture development, standardization, evaluation, and evolution across organizations or domains. & Business drivers, technology trends, stakeholder concerns, regulatory constraints, quality attribute requirements, and existing architectures. & Comprehensive architectural methodology, viewpoints, lifecycle processes \\ \hline
    \end{tabular}
    \caption{Comparative Overview of Architectural Concepts using the GIO Model}
    \label{tab:gio-comparison}
\end{table}

The comparison reveals that \textbf{Software Architecture} is the most concrete, focusing on specific system designs. \textbf{Reference Architecture} serves as a higher-level blueprint that generalizes solutions across a domain. Finally, \textbf{Architectural Frameworks} provide overarching structures and methodologies applicable across a wide range of domains and systems.


\section{Evaluating Abstract Architectures: Motivations and Challenges}
\label{sec:evaluating-abstract-architectures}

Architectural frameworks and reference models play a pivotal role in shaping software systems, particularly when dealing with large, complex families of systems. As the abstraction level increases, the challenges associated with evaluating the effectiveness of these models become significantly more pronounced. Traditional evaluation methods that focus on specific system architectures---such as the Architecture Tradeoff Analysis Method (ATAM)---fail to capture the dynamic variability and long-term evolution inherent in frameworks and reference architectures. Evaluating these abstract architectural forms requires new techniques that not only consider system-specific concerns but also the broader, meta-level impacts across multiple instances, domains, and use cases.

\subsection{The Need for Tradeoff and Risk Evaluation at Higher Abstraction Levels}

The shift from evaluating concrete system architectures to abstract models like architectural frameworks or reference architectures introduces a range of complexities. Frameworks and reference models are intended to support the development of multiple, diverse systems within a given domain, introducing flexibility and scalability across families of systems. However, the inherent variability in these models means that evaluating tradeoffs and risks must account for more than just the current system design; it must also consider how well these architectural forms will perform across various potential instantiations.

For instance, while a concrete system architecture might prioritize performance and security in its specific context, an architectural framework must provide the necessary flexibility to support multiple performance configurations, evolving security requirements, and adaptable scalability mechanisms. Thus, the evaluation must balance these attributes across a variety of potential system instantiations, ensuring that the framework supports a broad range of use cases without compromising key quality attributes.

\subsection{Limitations of Existing Methods (ATAM, ATAM/R)}

While the Architecture Tradeoff Analysis Method (ATAM) effectively evaluates concrete system architectures, it does not address reference architectures. To fill this gap, ATAM/R was developed, extending ATAM to handle the evaluation of reference architectures by considering multiple system instantiations based on a given template. However, while ATAM/R is suited for reference architectures, it falls short when applied to architectural frameworks, which provide a meta-level approach to designing families of systems with significant flexibility and variability.

Architectural frameworks introduce complexities that neither ATAM nor ATAM/R were designed to handle. Frameworks require evaluation methods that account for their adaptability, long-term evolution, and the diverse system instantiations they support. Thus, while ATAM/R addresses the concerns of reference architectures, it fails to capture the full breadth of considerations necessary for evaluating architectural frameworks, such as extensibility, flexibility, and lifecycle management.

\subsection{Requirements for a Multi-Level Evaluation Framework}

To address the limitations of current methods, a multi-level evaluation framework is needed---one that can adapt to both the static qualities of specific systems and the dynamic variability of frameworks and reference architectures. Such a framework must consider several key factors:

\begin{itemize}
    \item \textbf{Variability Management}: Architectural frameworks and reference architectures both support a range of possible system designs through variability mechanisms. However, architectural frameworks provide more advanced mechanisms for managing variability, including the explicit definition of \textit{extension points} that allow for significant customization and adaptation. These extension points are not a typical feature of reference architectures, which focus on domain-specific patterns and reuse but do not generally allow for the same level of flexibility.
    
    \item \textbf{Meta-Level Quality Attributes}: Unlike concrete system architectures, frameworks and reference architectures are primarily evaluated for their ability to support system families and long-term evolution. However, the nature of the meta-level quality attributes like flexibility, generalizability, and adaptability differs between reference architectures and architectural frameworks. For reference architectures, these qualities are generally domain-specific and focus on the ability to support different system instantiations within a specific domain. In contrast, for architectural frameworks, these qualities take on a broader, more meta-level role, emphasizing the ability to support a wide range of systems across diverse domains and long-term evolution.
    
    \item \textbf{Scenario Diversity}: The evaluation of scenarios varies significantly between software architectures, reference architectures, and architectural frameworks. For \textit{software architectures}, scenarios tend to be concrete and tied to specific system operations, focusing on functional, performance, and failure scenarios. In contrast, \textit{reference architectures} introduce generalized scenarios that reflect the variability across multiple system instantiations within a domain, such as adoption or interoperability. Finally, \textit{architectural frameworks} require scenarios that consider not only the variability across system families but also scenarios related to framework-level concerns like evolution, integration, and lifecycle support across diverse domains and contexts. These differences highlight the increasing level of abstraction and the corresponding complexity in managing scenarios at each level.
    
    \item \textbf{Stakeholder Involvement}: Stakeholder concerns differ across software architectures, reference architectures, and architectural frameworks. \textit{Software architecture} primarily involves designers, developers, and users focused on system-specific requirements. \textit{Reference architectures} bring in domain experts, architects, and organizations seeking consistency and reuse within a domain. \textit{Architectural frameworks}, however, involve a wider range of stakeholders, including process engineers, lifecycle managers, and decision-makers, all concerned with long-term adaptability, evolution, and support for multiple system families across domains.
\end{itemize}

A multi-level evaluation framework, like the proposed Architecture Tradeoff and Risk Analysis Framework (ATRAF), must address these needs by extending methods such as ATAM to handle the complexities of frameworks and reference architectures. ATRAF offers a comprehensive solution by integrating flexibility management, meta-level quality attributes, and multi-context scenario realization, providing the necessary tools to evaluate both concrete system architectures and abstract architectural models. It consists of three methods: ATRAM, which evaluates concrete software architectures; RATRAM, which adapts this for reference architectures, focusing on variability and reuse; and AFTRAM, which addresses the meta-level concerns of architectural frameworks, emphasizing extensibility, adaptability, and the long-term evolution of system families across diverse domains.

The following sections introduce the core principles of ATRAF, illustrating how its methods---ATRAM, RATRAM, and AFTRAM---extend traditional evaluation techniques to effectively address the needs of both concrete system architectures and more abstract architectural models.


\section{Methodology}
\label{sec:methodology}

\begin{figure}[ht]
    \centering
    \includegraphics[width=\textwidth]{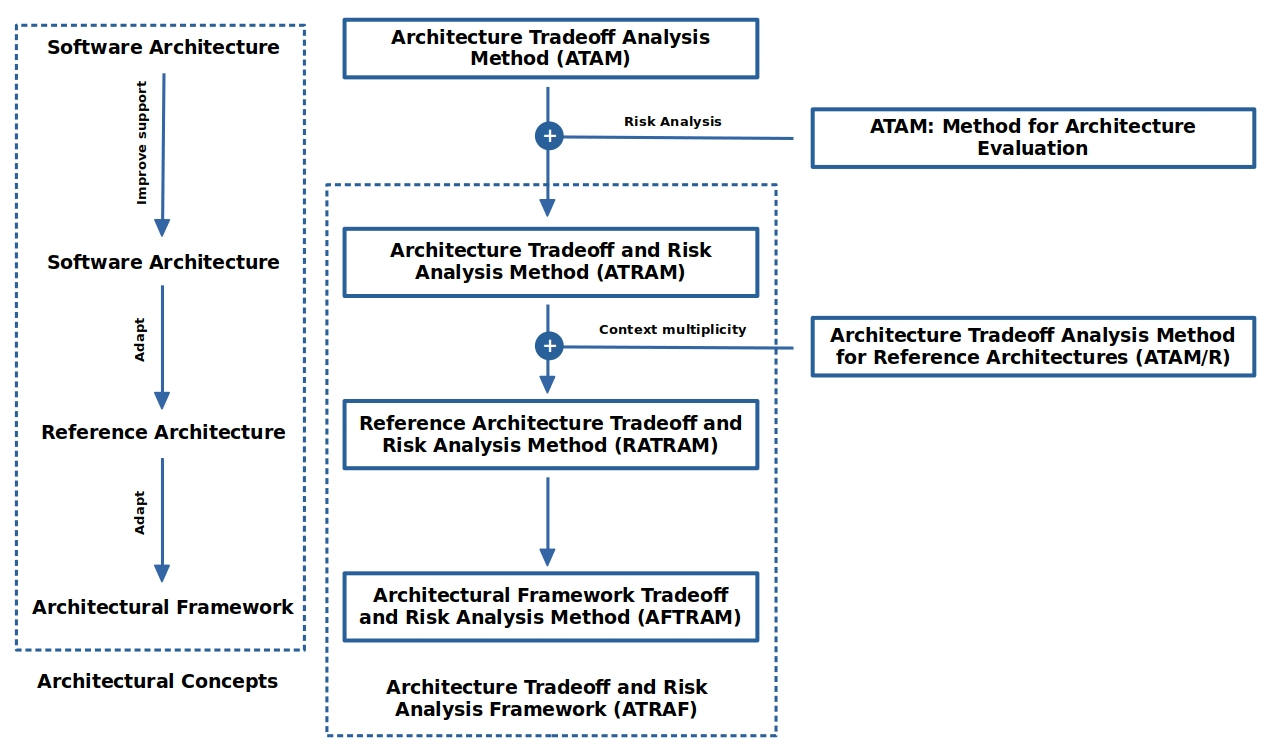}
    \caption{Derivation process of ATRAF}
    \label{fig:atraf-derivation-process}
\end{figure}

To address the limitations of existing architecture evaluation methods at higher abstraction levels, we developed the \textbf{Architecture Tradeoff and Risk Analysis Framework (ATRAF)}. ATRAF extends and unifies tradeoff and risk analysis across software architectures, reference architectures, and architectural frameworks. This section outlines the theoretical grounding, method derivation process, and iterative evaluation model that guided ATRAF’s construction (see Figure~\ref{fig:atraf-derivation-process}).

\subsection{Design Goals, Theoretical Foundations, and Related Methods}

ATRAF was designed to achieve three central goals. First, it aims to preserve the strengths of the original Architecture Tradeoff Analysis Method (ATAM), including its scenario-driven structure and its ability to reveal tradeoffs between competing quality attributes. Second, ATRAF explicitly supports risk identification, both as an analytical deliverable and as a driver of architecture refinement. Third, it extends the applicability of ATAM to more abstract architectural concepts---such as reference architectures and architectural frameworks---without sacrificing methodological rigor or traceability.

Our approach is conceptually rooted in the 1998 version of ATAM \cite{kazman1998atam}, which organizes the method into four explicit phases within a spiral model, emphasizing iterative refinement and continuous architectural improvement. This phased organization provides a solid pedagogical and analytical foundation, particularly valuable when adapting the method to abstract architectural forms like reference architectures and frameworks. While the 2000 ATAM report \cite{kazman2000atam} formalized a more prescriptive and linear 9-step process---well-suited for repeatable industrial applications---it places less emphasis on the iterative and evolutionary aspects of architectural design. In ATRAF, we deliberately retained the 1998 version's four-phase spiral structure to maintain its support for iteration and progressive refinement. We also integrated refinements from the 2000 report, such as more structured stakeholder workshops and clearer articulation of evaluation results (e.g., risk themes), which enhanced procedural clarity without diminishing the method's iterative spirit.

While ATAM and its extension ATAM/R provided valuable insights for evaluating system-specific and reference architectures, ATRAF formalizes and extends these concepts into three concrete methods, each tailored to the specific needs and semantics of the target abstraction level---software architecture, reference architecture, and architectural framework.

\subsection{Derivation Process: From ATAM to ATRAM, RATRAM, and AFTRAM}

The development of ATRAF involved systematically adapting the original Architecture Tradeoff Analysis Method (ATAM) to address the needs of higher abstraction levels, such as reference architectures and architectural frameworks. The method evolved through a series of deliberate extensions, drawing from foundational sources and progressively expanding on the methodology to accommodate more complex architectural evaluation scenarios.

\begin{itemize}
    \item \textbf{ATRAM (Architecture Tradeoff and Risk Analysis Method)}: ATRAM is based on the Architecture Tradeoff Analysis Method (ATAM), initially introduced by \cite{kazman1998atam} (1998). ATAM organizes the evaluation of software architectures into four iterative phases, promoting an incremental and feedback-driven approach to architecture improvement. ATRAM retains the four-phase structure of ATAM, aligning closely with the original model while formally incorporating risk identification, which was more explicitly emphasized in the 2000 ATAM report \cite{kazman2000atam}. In this report, titled \textit{ATAM: Method for Architecture Evaluation}, risk identification was formalized as a distinct activity within the evaluation process. ATRAM incorporates this risk identification step, allowing architecture teams to identify potential risks and refine their design decisions early in the evaluation process.

    \item \textbf{RATRAM (Reference Architecture Tradeoff and Risk Analysis Method)}: RATRAM builds upon ATRAM, adapting the method for the evaluation of reference architectures. It incorporates key insights from the \textit{Architecture Tradeoff Analysis Method for Reference Architectures (ATAM/R)}\cite{angelov2014atamr}. While RATRAM retains the core principles of ATRAM, it introduces several key innovations, including the concept of \textit{context multiplicity,} which evaluates architectures across multiple domains and deployment contexts. Furthermore, RATRAM draws on ATAM/R’s approach to handling aggregated architectures, enabling a more comprehensive analysis of complex reference architectures. In addition, RATRAM introduces domain-aligned scenario types---such as adoption, interoperability, and scalability scenarios---along with stakeholder models specifically tailored to reference architectures. These adaptations allow RATRAM to assess systems across a broader range of reuse and context variability, offering greater flexibility in managing architectural tradeoffs and better supporting decision-making in diverse contexts.

    \item \textbf{AFTRAM (Architectural Framework Tradeoff and Risk Analysis Method)}: AFTRAM extends RATRAM by focusing on architectural frameworks, which serve as meta-architectures for designing system families. AFTRAM introduces advanced concepts such as meta-scenario simulation, enabling designers to assess the impact of architectural decisions across various system family configurations. Additionally, it incorporates process viewpoint modeling to evaluate lifecycle-wide tradeoffs and the evolution of architectural frameworks. These innovations further differentiate AFTRAM from RATRAM, providing a comprehensive evaluation approach for frameworks that must balance flexibility, standardization, and maintainability over time.

\end{itemize}

These methods---ATRAM, RATRAM, and AFTRAM---each build upon the iterative, four-phase framework of ATAM (1998) while incorporating later insights from the ATAM technical reports \cite{kazman2000atam} \cite{angelov2014atamr} to address evolving needs in architectural evaluation. The figure below provides a visual summary of the development process, illustrating how ATRAF emerged as a comprehensive framework for evaluating software architectures, reference architectures, and architectural frameworks.

\subsection{Validation Strategy Using the RTS Case Family}

To validate the applicability of ATRAF, we used a progressive example chain derived from the Remote Temperature Sensor (RTS) case, crafted specifically for illustrative purposes. This case family provides examples at each level of architectural abstraction to demonstrate how ATRAF’s methods apply to concrete system architectures, reference architectures, and architectural frameworks. The case family includes: (1) \textbf{Remote Temperature Sensor Architecture (RTSA)} (Appendix~\ref{app:rtsa-description}), a concrete system architecture evaluated using ATRAM; (2) \textbf{Remote Temperature System Reference Architecture (RTSRA)} (Appendix~\ref{app:rtsra-description}), a domain-level reference architecture evaluated using RATRAM; and (3) \textbf{Remote Monitoring Architectural Framework (RMAF)} (Appendix~\ref{app:rmaf-description}), an architectural framework that guides the design of monitoring systems across various domains, evaluated using AFTRAM. This vertical alignment of examples allows us to test ATRAF’s principles across the three levels of abstraction, ensuring consistency, effectiveness, and adaptability in the evaluation process.

The following sections of this paper will explore the details of each method within the Architecture Tradeoff and Risk Analysis Framework (ATRAF). We will begin with the Architecture Tradeoff and Risk Analysis Method (ATRAM)~\ref{sec:atram}, followed by the Reference Architecture Tradeoff and Risk Analysis Method (RATRAM)~\ref{sec:ratram}, and conclude with the Architectural Framework Tradeoff and Risk Analysis Method (AFTRAM)~\ref{sec:aftram}.


\section{ATRAM: Architecture Tradeoff and Risk Analysis Method}
\label{sec:atram}

The Architecture Tradeoff and Risk Analysis Method (ATRAM) extends the foundational work of the Architecture Tradeoff Analysis Method (ATAM), providing a structured and comprehensive approach to evaluating software architectures. While ATAM, introduced in 1998, emphasized an iterative spiral process that facilitated continuous architectural refinement, the 2000 refinement of ATAM added key enhancements such as improved risk identification and more explicit stakeholder involvement, offering greater procedural clarity.

ATRAM builds on the original principles of ATAM, incorporating these 2000 improvements while retaining the flexibility of the original spiral process, which supports ongoing, iterative evaluation and refinement of system architectures. This iterative approach allows for the continuous re-assessment of architectural decisions based on stakeholder feedback, evolving requirements, and the identification of risks as the system design matures.

A key strength of ATRAM is its ability to balance competing quality attributes such as performance, modifiability, and security. By integrating risk management into the evaluation process, ATRAM helps identify potential vulnerabilities and areas of uncertainty early, guiding decision-making throughout the system lifecycle. It provides architects with the tools needed to navigate complex design tradeoffs while ensuring the architecture meets both technical and stakeholder needs.

In this section, we explore the key evaluation phases of ATRAM, detailing how each phase contributes to identifying sensitivities, tradeoffs, and risks, and ultimately supports the refinement of software architectures.

\begin{figure}[ht]
    \centering
    \includegraphics[width=\textwidth]{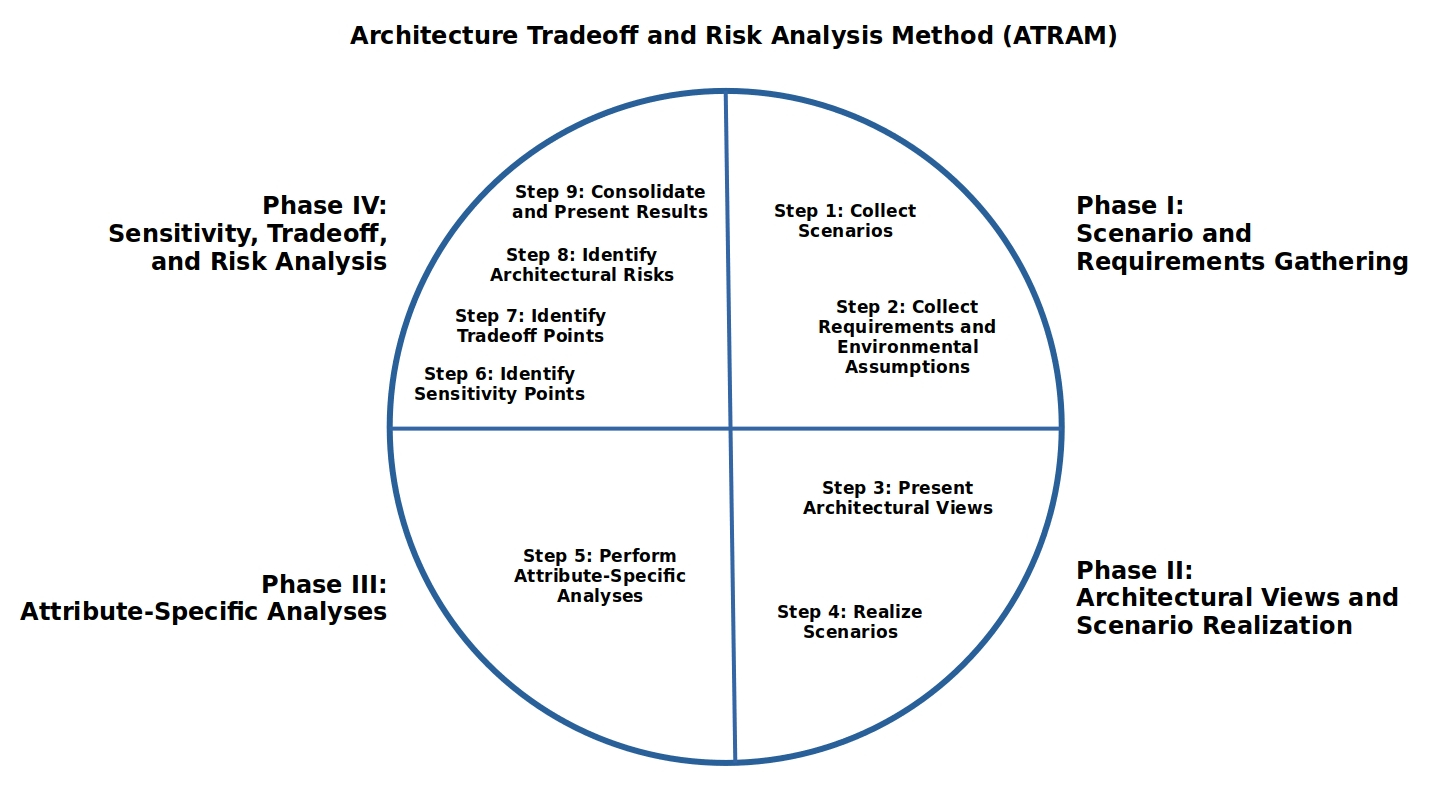}
    \caption{Steps of the Architecture Tradeoff and Risk Analysis Method (ATRAM)}
    \label{fig:atram-spiral-model}
\end{figure}

\subsection{Evaluation Phases}

The ATRAM method organizes the evaluation process into four distinct phases: \textbf{Phase I: Scenario and Requirements Gathering}, where relevant scenarios and system requirements are collected; \textbf{Phase II: Architectural Views and Scenario Realization}, in which the architecture is presented through various viewpoints and scenarios are mapped to assess architectural support; \textbf{Phase III: Attribute-Specific Analyses}, focusing on the evaluation of quality attributes such as performance and security; and \textbf{Phase IV:  Sensitivity, Tradeoff, and Risk Analysis}, which synthesizes findings to identify sensitivities, tradeoffs, and risks, guiding architectural refinements. These four phases collectively comprise 9 steps, as depicted in the spiral model shown in Figure~\ref{fig:atram-spiral-model}, illustrating the iterative nature of the evaluation process.

\subsubsection{Phase I: Scenario and Requirements Gathering}

\textbf{Purpose:}
To establish a structured foundation for evaluating a specific software architecture by systematically capturing stakeholder needs, system requirements, and operational scenarios. Scenario elicitation and requirements gathering are intertwined and iterative.

\textbf{Step 1: Collect Scenarios}
\begin{itemize}
    \item \textbf{Elicit scenarios from stakeholders}:
        \begin{itemize}
            \item Functional scenarios (real-world usage of the system)
            \item Evolution scenarios (anticipated system changes or enhancements)
            \item Stress scenarios (extreme or failure conditions)
        \end{itemize}
    \item \textbf{Facilitate collaborative workshops} to brainstorm scenarios and gather diverse concerns.
    \item \textbf{Prioritize scenarios} through stakeholder voting based on importance and perceived risk.
\end{itemize}

\textbf{Stakeholders}
\begin{itemize}
    \item Software architects (designers)
    \item Developers (implementers)
    \item Domain experts (e.g., security, QA)
    \item System operators and users (operational roles)
\end{itemize}

\textbf{Step 2: Collect Requirements and Environmental Assumptions}
\begin{itemize}
    \item Capture business drivers (mission goals, constraints, time-to-market, etc.)
    \item Elicit quality attribute requirements (performance, security, availability, etc.)
    \item Document technical constraints and platform assumptions
\end{itemize}

\textbf{Phase I Artifacts:}

\begin{tabular}{|p{4cm}|p{6cm}|p{5cm}|}
\hline  
Artifact & Description & Creation Process \\ \hline  
Stakeholder Map & Identifies stakeholder roles and their influence on decisions. & Interviews and workshops \\ \hline  
Scenario Catalog & Functional, evolution, and stress scenarios for analysis. & Brainstorming and interviews \\ \hline  
Prioritized Scenario List & Top scenarios ranked by risk and importance. & Stakeholder consensus \\ \hline  
System Requirements Document & Captures key requirements and business drivers. & Joint creation with stakeholders \\ \hline  
Environmental Assumptions Document & Technical and environmental assumptions documented. & Elicited in context discussions \\ \hline  
Utility Tree & Quality attributes $\rightarrow$ sub-attributes $\rightarrow$ scenarios, annotated with risk/importance. & Built collaboratively \\ \hline  
\end{tabular}

\subsubsection{Phase II: Architectural Views and Scenario Realization}

\textit{Purpose:}
To present the architecture in multiple views and trace how it supports high-priority scenarios. This provides the structural foundation for subsequent evaluations of how well the architecture supports desired quality attributes.

\textbf{Step 3: Present Architectural Views}

\textit{Includes identification of architectural tactics/approaches}

\textbf{Description}
\begin{itemize}
    \item Present the software architecture using the following architectural viewpoints:
        \begin{itemize}
            \item Structural Viewpoint: modules and components, their responsibilities, and relationships within the software architecture
            \item Interaction Viewpoint: data exchange and communication paths between components
            \item Behavioral Viewpoint: runtime behavior, control flow, and concurrency strategies
            \item Deployment Viewpoint: the distribution of software components across the execution environment (e.g., hardware nodes, containers)
        \end{itemize}
    \item Identify the architectural styles, design patterns, and tactics employed to support key quality attributes such as performance, availability, and modifiability.
\end{itemize}

\textbf{Artifacts:}

\begin{tabular}{|p{4cm}|p{6cm}|p{5cm}|}
\hline  
Artifact & Description & Creation Process \\ \hline  
Architectural Viewpoints Document & Viewpoints of the system showing structure, behavior, and deployment. & Prepared by architecture team using standard modeling practices. \\ \hline  
Architectural Approaches & Design patterns, styles, or tactics used to support quality attributes. & Identified collaboratively with architects; document rationale for each. \\ \hline  
\end{tabular}

\textbf{Step 4: Realize Scenarios}

\begin{itemize}
    \item \textbf{For each high-priority scenario}, analyze how the software architecture enables or constrains its realization.
    \item \textbf{Describe the realization strategy}:
        \begin{itemize}
            \item Identify which architectural elements (e.g., components, connectors, modules) are involved in the scenario’s realization
            \item Trace the flow of control and data through the architecture
            \item Document any design assumptions made regarding behavior, environment, or usage context
            \item Clarify whether the scenario can be structurally realized without modification, or whether it exposes architectural tensions
        \end{itemize}
    \item \textbf{Classify scenario support level}:
        \begin{itemize}
            \item Natively Supported---the scenario is clearly and directly supported by the existing architecture
            \item Constrained Realization---the architecture supports the scenario only under certain assumptions or limitations
            \item Unsupported---structural misalignments prevent realistic realization of the scenario
        \end{itemize}
    \item \textbf{Identify realization gaps}:
        \begin{itemize}
            \item Describe any architectural areas where scenario needs are only partially met
            \item Highlight structural limitations or inflexibilities
            \item Suggest refinements or architectural adjustments that could enable improved support
        \end{itemize}
\end{itemize}

\textbf{Artifacts:}

\begin{tabular}{|p{4cm}|p{6cm}|p{5cm}|}
\hline  
Artifact & Description & Creation Process \\ \hline  
Scenario Realization Document & Architecture-specific realizations of each scenario, showing flow/control/data paths. Mapping of scenario elements to architectural constructs. Design and behavior assumptions made during scenario mapping. & Created by tracing scenarios through components and connectors. Captured interactively during workshops. \\ \hline  
\end{tabular}

\subsubsection{Phase III: Attribute-Specific Analyses}

\textit{Purpose:}
To analyze each quality attribute of the software architecture in isolation, based on the architectural views and scenario realizations. This phase provides attribute-specific insight without identifying tradeoffs, sensitivities, or risks. It supports separation of concerns by allowing quality attributes to be evaluated independently, enabling domain experts to contribute targeted analysis using appropriate techniques.  

\textbf{Step 5: Perform Attribute-Specific Analyses}

Each prioritized quality attribute---such as performance, modifiability, or availability---is evaluated separately with respect to the software architecture. Analyses may use informal (expert reasoning), semi-formal (scenario-based questioning), or formal (quantitative modeling) methods. Evaluations focus on estimating attribute-related behaviors (e.g., response times, change effort, fault tolerance) using the architectural views and scenario mappings from Phase II. No cross-attribute critique or tradeoff analysis is performed at this point.

\textbf{Artifacts:}

\begin{tabular}{|p{4cm}|p{6cm}|p{5cm}|}
\hline  
Artifact & Description & Creation Process \\ \hline  
Attribute Evaluation Report & For each attribute, documents the evaluation rationale, analysis method, assumptions, estimated behavior, and relevant observations regarding how the software architecture supports or limits the attribute. & Developed by analysts or domain experts based on scenarios and views from Phase II. \\ \hline  
\end{tabular}

\subsubsection{Phase IV: Sensitivity, Tradeoff, and Risk Analysis}

\textit{Purpose:}
To interpret the results of attribute-specific analyses and transform them into actionable architectural insights. This phase identifies areas of architectural sensitivity, reveals tradeoffs between competing attributes, highlights architectural risks, and consolidates findings to guide stakeholder decision-making and architecture refinement.

\textbf{Step 6: Identify Sensitivity Points}

Determine which architectural elements have a strong influence on a single quality attribute. These are sensitivity points---areas where small changes may cause significant effects. Understanding them highlights key leverage points in the architecture and helps anticipate consequences of design evolution.

\textbf{Artifact:}

\begin{tabular}{|p{4cm}|p{6cm}|p{5cm}|}
\hline  
Artifact & Description & Creation Process \\ \hline  
Sensitivity Point List & List of sensitive architectural elements, each annotated with: (1) associated attribute, (2) impact type (e.g., latency, cost), and (3) key assumptions made during analysis. & Derived from attribute-specific models (Phase III) and assumption logs (Phases II–III). \\ \hline  
\end{tabular}  

\textbf{Step 7: Identify Tradeoff Points}

Identify architectural elements that simultaneously affect multiple quality attributes, often in conflicting ways. These are tradeoff points---locations where improvement in one attribute may degrade another. Use high-priority scenarios (from Phase I) and attribute analyses (from Phase III) to trace where such tensions arise. Clarify the nature of each tradeoff and how it manifests in the architectural structure or behavior.

\textbf{Artifact:}

\begin{tabular}{|p{4cm}|p{6cm}|p{5cm}|}  
\hline  
Artifact & Description & Creation Process \\ \hline  
Tradeoff Point Matrix & Table showing architectural elements that influence two or more attributes, annotated with the direction of influence (e.g., ++ performance, -- security), and linked to scenarios that highlight the tradeoff. & Created by cross-referencing the Sensitivity Point List with scenario traces and attribute outcomes. \\ \hline  
\end{tabular}

\textbf{Step 8: Identify Architectural Risks}

Document architectural risks---areas of uncertainty, fragility, or deferred decisions. Risks may stem from unsupported assumptions, unresolved requirements, incomplete data, or elements prone to failure or misinterpretation. While some risks overlap with sensitivities and tradeoffs, others arise independently and must be captured for resolution planning.

\textbf{Artifact:}

\begin{tabular}{|p{4cm}|p{6cm}|p{5cm}|}
\hline  
Artifact & Description & Creation Process \\ \hline  
Architectural Risk Document & Structured list of risks with metadata: (1) affected elements, (2) related attribute(s), (3) risk origin (e.g., assumption, missing input), and (4) projected impact. & Synthesized from evaluation notes and validated collaboratively with stakeholders. \\ \hline  
\end{tabular}

\textbf{Step 9: Consolidate and Present Results}

Synthesize the findings from prior steps into a clear, stakeholder-facing summary. This includes key sensitivities, critical tradeoffs, and unresolved risks, along with practical recommendations. The goal is to equip decision-makers with a comprehensive understanding of architectural implications and help guide next steps such as redesign, refinement, or further analysis.

\textbf{Artifacts:}

\begin{tabular}{|p{4cm}|p{6cm}|p{5cm}|}
\hline  
Artifact & Description & Creation Process \\ \hline  
Evaluation Summary Report & Integrated document presenting traceable findings, analysis rationale, and prioritized recommendations for action. & Compiled by evaluation team as a final deliverable. \\ \hline  
Action Plan Recommendations & Clear list of follow-up actions including architectural changes, risk mitigations, stakeholder clarifications, or prototypes. & Developed collaboratively in final stakeholder session. \\ \hline  
\end{tabular}

\subsubsection{Iteration and Refinement in ATRAM}

ATRAM is structured as a spiral process, where each phase informs and may trigger refinements in earlier phases. If attribute-specific analyses or tradeoff identification reveal significant gaps---such as unmet quality goals, unexpected risks, or structural tensions---an action plan is developed to adjust the architecture, revise assumptions, or refine requirements. This leads naturally to another iteration of the method, enabling progressive convergence toward a viable design. While the method is presented in discrete steps, in practice these phases interact fluidly: architectural modeling may expose requirement ambiguities, and scenario analysis may reshape stakeholder priorities. ATRAM encourages continuous analysis throughout the system lifecycle, supporting architecture evolution from design through deployment and maintenance.

\subsection{Case Study: Remote Temperature Sensor Architecture (RTSA)}

To validate the applicability of ATRAM, the Remote Temperature Sensor Architecture (RTSA) was specifically designed as an example to evaluate ATRAM's application to concrete software architectures. The example describes a system for monitoring furnace temperatures, detailing its three main components, operational flows, and security considerations. (For a more comprehensive description of the RTSA, see Appendix~\ref{app:rtsa-description}).

\noindent \textbf{Note:} This is an ongoing work. Future versions of this paper will incorporate extensive case-specific evaluations of the RTSA using the Architecture Tradeoff Analysis Method (ATRAM).


\section{RATRAM: Reference Architecture Tradeoff and Risk Analysis Method}
\label{sec:ratram}

Reference architectures capture domain-wide reusable architectural patterns, styles, and guidelines that can be instantiated into multiple system-specific software architectures. While not tied to a single implementation, they directly shape derived architectures' quality attributes and evolution. Unlike architectural frameworks (which are meta-architectures) or software architectures (which are system-specific), reference architectures operate at a middle abstraction level: general yet actionable.

Because of this position, reference architectures impose domain-wide design constraints, support consistency, and promote best practices --- but may also embed tradeoffs, structural inflexibilities, or implicit risks. Existing architecture evaluation methods like ATAM and ATRAM focus on concrete software systems. To evaluate reference architectures’ structural and quality attribute implications, we introduce the \textbf{Reference Architecture Tradeoff and Risk Analysis Method (RATRAM)}.

RATRAM is an adaptation of ATRAM tailored to the domain-level generality of reference architectures, incorporating both well-structured and aggregated forms, aspects borrowed from ATAM/R \cite{angelov2014atamr} as detailed in our Methodology, Section~\ref{sec:methodology}.

\begin{figure}[ht]
    \centering
    \includegraphics[width=\textwidth]{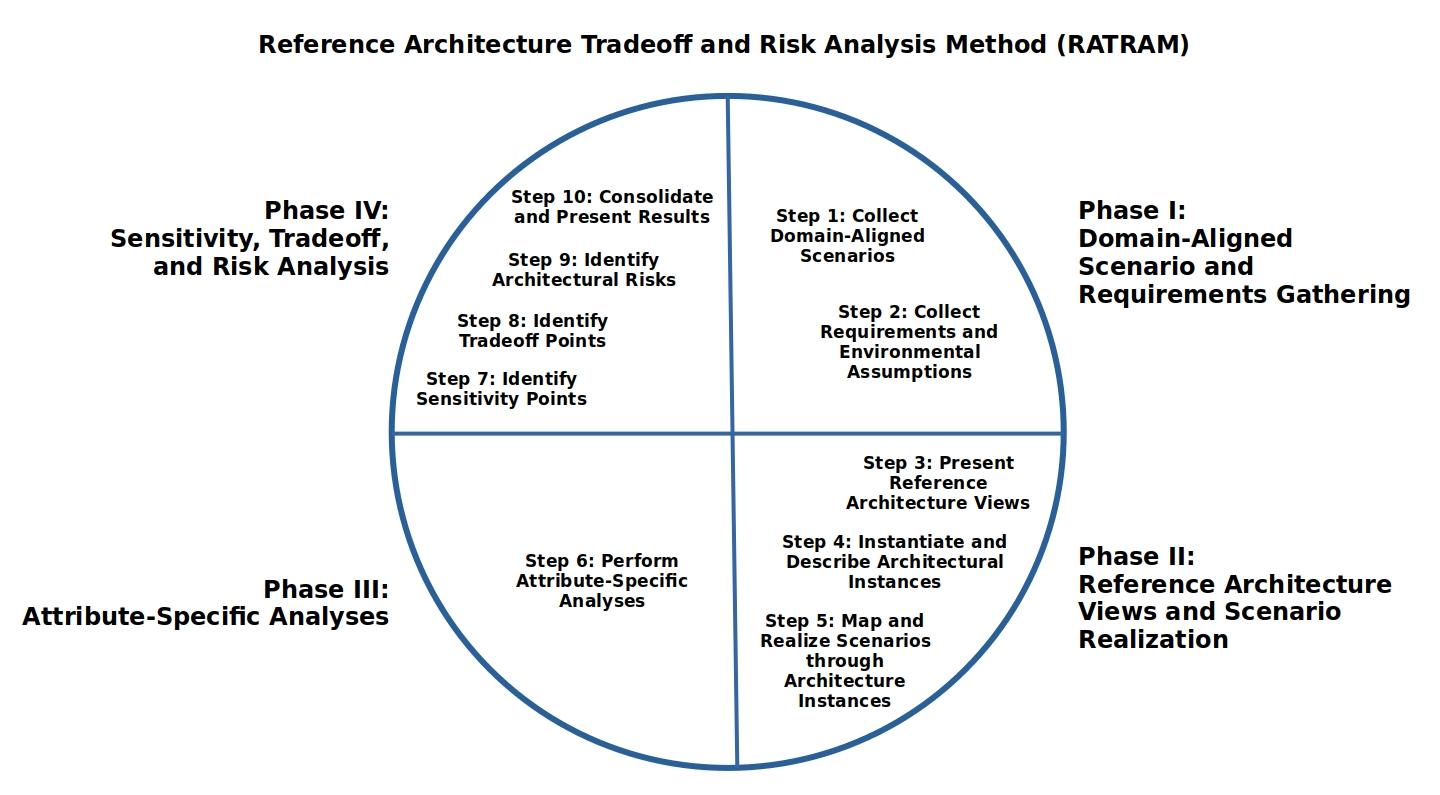}
    \caption{Steps of the Reference Architecture Tradeoff and Risk Analysis Method (RATRAM)}
    \label{fig:ratram-spiral-model}
\end{figure}

\subsection{Evaluation Phases}

Similarly to ATRAM, the RATRAM method organizes its evaluation process into four phases with carefully adjusted artifacts and evaluation targets, along with minor modifications in the naming of Phase I and II to emphasize domain-specific alignment: \textbf{Phase I: Domain-Aligned Scenario and Requirements Gathering}, where domain-relevant scenarios and system requirements are systematically identified and documented; \textbf{Phase II: Reference Architecture Views and Scenario Realization}, which involves representing the architecture through multiple viewpoints and mapping scenarios to assess its structural and functional adequacy; \textbf{Phase III: Attribute-Specific Analyses}, which evaluates individual quality attributes such as performance and security within the intended domain context; and \textbf{Phase IV: Sensitivity, Tradeoff, and Risk Analysis}, synthesizing findings to identify critical sensitivities, trade-offs, and risks, thereby guiding architectural refinements. Collectively, these four phases encompass 10 iterative steps, as depicted in the spiral model shown in Figure~\ref{fig:ratram-spiral-model}, underscoring the cyclical and dynamic aspects of the evaluation process.

\subsubsection{Phase I: Domain-Aligned Scenario and Requirements Gathering}

\textbf{Purpose:}
To build a structured foundation for evaluating a \textbf{Reference Architecture (RA)} by capturing domain-level concerns, reusable architectural goals, and patterns. This includes eliciting how the RA is used, instantiated, evolved, and challenged in real-world domain contexts. Scenario elicitation and requirements gathering are intertwined and iterative.

\textbf{Step 1: Collect Domain-Aligned Scenarios}
\begin{itemize}
    \item \textbf{Elicit generalized scenarios}:
    \begin{itemize}
        \item Functional scenarios (canonical usage across systems in the domain)
        \item Evolution scenarios (domain change, new technologies)
        \item Stress scenarios (boundary and challenge conditions)
        \item Adoption scenarios (instantiation by system architects)
        \item Interoperability Scenarios
    \end{itemize}
\end{itemize}

\textbf{Stakeholders:}
\begin{itemize}
    \item RA designers (designers)
    \item System architects (implementers of derived systems)
    \item Domain experts and standards authorities (domain experts)
    \item Operational leads from applying organizations (operational roles)
\end{itemize}

\textbf{Step 2: Collect Requirements and Environmental Assumptions}
\begin{itemize}
    \item \textbf{Reference-Level Requirements:}
    \begin{itemize}
        \item Architectural patterns, instantiation mechanisms, reuse expectations, modularity
    \end{itemize}

    \item \textbf{Derived-System Requirements:}
    \begin{itemize}
        \item Functional and quality capabilities enabled in downstream systems
        \item Traceability to domain-wide goals
    \end{itemize}

    \item \textbf{Environmental Assumptions:}
    \begin{itemize}
        \item Regulatory conditions, domain-specific technology constraints, lifecycle models
    \end{itemize}
\end{itemize}

\textbf{Phase I Artifacts:}

\begin{tabular}{|p{4cm}|p{6cm}|p{5cm}|}
\hline
Artifact & Description & Creation Process \\
\hline
Stakeholder \& Domain Map & Identifies RA stakeholders and domain forces shaping the architecture. & Workshops, interviews \\ \hline  
Scenario Catalog & Generalized functional, evolution, stress, adoption, and quality attribute scenarios. & Domain-focused elicitation \\ \hline  
Prioritized Scenario List & High-impact scenarios for deeper analysis. & Risk/importance voting \\ \hline  
Reference-Level Requirement Document & RA structure, reuse, and quality goals. & Synthesized from stakeholders \\ \hline  
Derived-System Requirement Document & System needs RA must support. & Collected with traceability mapping \\ \hline  
Environmental Assumptions Document & Domain and platform constraints, assumptions. & Captured in joint sessions \\ \hline  
Utility Tree & Quality attribute breakdown guiding focus. & Built collaboratively \\
\hline
\end{tabular}

\subsubsection{Phase II: Reference Architecture Views and Scenario Realization}

\textbf{Purpose:}
To represent the structure and rationale of the reference architecture and analyze how it supports prioritized domain scenarios.

\textbf{Step 3: Present Reference Architecture Views}

\begin{itemize}
    \item \textbf{Description:}
    \begin{itemize}
        \item Represent the reference architecture using architectural viewpoints appropriate for capturing domain-level structures:
        \begin{itemize}
            \item \textbf{Structural Viewpoint}: abstract modules and component roles, along with their responsibilities and interrelationships
            \item \textbf{Interaction Viewpoint}: communication paths and coordination patterns between roles or system entities
            \item \textbf{Behavioral Viewpoint}: intended runtime dynamics of common component interactions within systems derived from the reference architecture
            \item \textbf{Variability Viewpoint}: designated variation points, allowed configuration strategies, and guidance for tailoring the architecture to specific systems
        \end{itemize}
        \item Define the architectural patterns, templates, and integration mechanisms provided by the reference architecture.
        \item Describe the rationale behind major structural or interaction patterns, especially where they promote domain-wide quality attributes such as scalability, interoperability, or security.
        \item Specify any architectural constraints or design rules that must be respected during instantiation.
    \end{itemize}
\end{itemize}

\textbf{Artifacts:}

\begin{tabular}{|p{4cm}|p{6cm}|p{5cm}|}
\hline
Artifact & Description & Creation Process \\
\hline
Reference Architecture Viewpoints Document & Logical/structural representations showing domain-relevant abstractions using structural, interaction, and variability viewpoints. & Prepared by RA designers using standard modeling techniques. \\ \hline
Pattern and Constraint Map & Describes architectural rules, design patterns, and constraints imposed on implementers. & Extracted from RA specifications and refinement workshops. \\
\hline
\end{tabular}

\textbf{Step 4: Instantiate and Describe Architectural Instances}

\begin{itemize}
    \item \textbf{Instantiate one or more candidate system architectures} by leveraging the structural elements, architectural patterns, and any defined variation points of the reference architecture. Depending on its nature, the reference architecture may be a configurable architecture with explicit variability or an aggregated architecture composed of multiple predefined instances. The resulting candidate architectures should reflect the reference architecture’s applicability across different domain scenarios or operational contexts.

    \item \textbf{For each instantiated architecture:}
    \begin{itemize}
        \item \textbf{Describe the structure:}
        \begin{itemize}
            \item Identify the component configurations and their roles, derived from the reference architecture
            \item Indicate which elements are:
            \begin{itemize}
                \item Reused directly (i.e., adopted as-is from the reference architecture)
                \item Reused with customization via variation points (configured or extended as intended by the variability guidance)
                \item Extended or newly introduced (elements not anticipated by the reference architecture)
            \end{itemize}
        \end{itemize}
        \item \textbf{Capture instantiation details:}
        \begin{itemize}
            \item Outline the architectural constraints followed or relaxed
            \item Document any assumptions or design decisions made during instantiation
            \item Highlight the variation points exercised and the customization strategies applied
        \end{itemize}

        \item \textbf{Map the instantiation to architectural mechanisms:}
        \begin{itemize}
            \item Clarify how the instantiated elements trace back to templates, patterns, or constraints provided by the reference architecture
            \item Identify the mechanisms used to support reuse and adaptation, such as modular patterns, interface contracts, or substitution guidelines
        \end{itemize}

        \item \textbf{Motivate the instantiated architecture:}
        \begin{itemize}
            \item Specify which domain scenarios the architecture is designed to support
            \item Argue for its suitability to address those scenarios and quality attribute goals, based on the structure and rationale of the reference architecture
            \item Describe tradeoffs considered, such as alignment with domain-specific requirements versus flexibility or scalability
            \item Note any structural mismatches or challenges encountered during instantiation
        \end{itemize}
    \end{itemize}
\end{itemize}

\textbf{Artifacts:}

\begin{tabular}{|p{4cm}|p{6cm}|p{5cm}|}
\hline
Artifact & Description & Creation Process \\
\hline
Architectural Instances Catalog & Document instantiated architectures derived from the reference architecture, including their structure, configuration decisions, motivations for scenario support. & Created through instantiation exercises and architectural analysis of RA reuse/adaptation. \\
\hline
\end{tabular}

\textbf{Step 5: Map and Realize Scenarios through Architecture Instances}

\begin{itemize}
    \item \textbf{For each scenario} from the Scenario Catalog, analyze how the reference architecture supports or constrains its realization within one or more of the instantiated system architectures.

    \item \textbf{Map each scenario to architectural instances} by:
    \begin{itemize}
        \item Identifying the architectural elements involved in the scenario's realization
        \item Tracing how those elements relate to the original reference architecture (as-is reuse, variation-based customization, or newly introduced)
        \item Highlighting how the instantiated components and interaction patterns respond to scenario requirements
    \end{itemize}

    \item \textbf{Describe the scenario realization strategy} for each case:
    \begin{itemize}
        \item Show how control and data flow through components
        \item Document any configuration, integration, or adaptation steps taken during realization
        \item Capture behavioral assumptions and environmental dependencies
    \end{itemize}

    \item \textbf{Classify the realization support level:}
    \begin{itemize}
        \item \textbf{Native Support} --- scenario is fully supported without modification
        \item \textbf{Guided Realization} --- realization follows clearly defined variation or adaptation guidance
        \item \textbf{Partial Realization} --- realization requires workaround or interpretation beyond the documented RA scope
        \item \textbf{Unsupported} --- realization is misaligned with or not feasible under current RA structure
    \end{itemize}

    \item \textbf{Identify realization gaps}, if applicable:
    \begin{itemize}
        \item Highlight parts of the scenario that lack sufficient architectural support or demand additional customization
        \item Point out structural tensions or misalignments between the RA and scenario demands
        \item Recommend refinements or extensions to the RA where appropriate
    \end{itemize}
\end{itemize}

\textbf{Artifacts:}

\begin{tabular}{|p{4cm}|p{6cm}|p{5cm}|}
\hline
Artifact & Description & Creation Process \\
\hline
Scenario Realization Document & Consolidates scenario-to-instance mappings, realization strategies, support classification (e.g., native, guided), and any identified gaps or misalignments between the reference architecture and scenario needs. & Produced through scenario walkthroughs using architecture models; documented collaboratively with stakeholders. \\
\hline
\end{tabular}

\subsubsection{Phase III: Attribute-Specific Analyses}

\textbf{Purpose:}
To evaluate how the reference architecture supports individual quality attributes \textbf{across its intended domain}, without analyzing attribute interactions or tradeoffs. This phase focuses on how architectural structures, patterns, or variation mechanisms influence attribute realization in derived systems. For aggregated architectures---those composed of consolidated design decisions---analysis emphasizes abstract properties such as scoping and applicability, reflecting their higher level of abstraction.

\textbf{Step 6: Perform Attribute-Specific Analyses}

Each selected quality attribute is analyzed independently using architectural viewpoints, scenario-driven instances, and structural guidance provided by the reference architecture. Evaluation methods include expert judgment, constraint analysis, or domain modeling. The goal is to estimate how well the reference architecture enables attribute realization across multiple instantiations, taking into account reuse strategies, abstraction constraints, and domain expectations.

\textbf{Artifacts:}

\begin{tabular}{|p{4cm}|p{6cm}|p{5cm}|}
\hline
Artifact & Description & Creation Process \\
\hline
Attribute Evaluation Report & Attribute-specific findings at the reference architecture level, including rationale, expected impact on derived systems, assumptions, and reuse or variability considerations. & Developed by domain analysts or evaluators using scenario and instance evidence from Phase II. \\
\hline
\end{tabular}

\subsubsection{Phase IV: Sensitivity, Tradeoff, and Risk Analysis}

RATRAM Phase IV follows the same structure and intent as ATRAM Phase IV, applying its four steps --- Identify Sensitivity Points, Identify Tradeoff Points, Identify Architectural Risks, and Consolidate and Present Results --- to the context of reference architectures. The key nuance lies in the domain-level abstraction of reference architectures: sensitivities and tradeoffs are often associated with reusable patterns, variation points, or architectural roles. In aggregated architectures, such concerns tend to surface at the level of styles and overarching design intent, rather than within fine-grained structural elements.

\subsubsection{Iteration and Refinement in RATRAM}

Like ATAM and ATRAM, RATRAM follows a spiral model that supports iterative refinement. Insights from attribute analyses, tradeoffs, or risks may lead to adjustments in reference architecture structures, variation mechanisms, or domain assumptions, prompting re-evaluation through earlier phases. This ensures that reference architectures remain adaptable and aligned with evolving domain demands while preserving structural integrity and reuse value.

\subsection{Case Study: Remote Temperature System Reference Architecture (RTSRA)}

To validate the applicability of RATRAM, the Remote Temperature System Reference Architecture (RTSRA) was specifically designed as an example to evaluate RATRAM's application to domain-level reference architectures. The RTSRA provides adaptable patterns and guidelines for remote temperature monitoring across various industries, detailing its core components, data acquisition protocols, fault tolerance mechanisms, and communication strategies. It supports diverse sensor types such as furnaces and air conditioning systems, while emphasizing flexibility, scalability, and customizable fault recovery strategies. (For a more comprehensive description of the RTSRA, see Appendix~\ref{app:rtsra-description}). 

\noindent \textbf{Note:} This is an ongoing work. Future versions of this paper will incorporate extensive case-specific evaluations of the RTSRA using the Reference Architecture Tradeoff Analysis Method (RATRAM).


\section{AFTRAM: Architectural Framework Tradeoff and Risk Analysis Method}
\label{sec:aftram}

Architectural frameworks are structured meta-architectures that provide methodologies, principles, and patterns to guide the design of families of related systems across diverse domains. They guide the development of these systems but inherently impose tradeoffs that shape key quality attributes such as modifiability, security, scalability, and performance. Existing evaluation techniques, including the Architecture Tradeoff Analysis Method (ATAM), focus on concrete system architectures and are not suited to the broader and more abstract nature of frameworks. The abstraction and variability inherent to frameworks require distinct evaluation approaches. To address this gap, we present the Architectural Framework Tradeoff and Risk Analysis Method (AFTRAM), a structured approach to assessing how frameworks influence the achievement of multiple, often competing, quality attributes. Enhancing one attribute often comes at the expense of others, and AFTRAM supports structured reasoning about these interdependencies. Similar to ATRAM, AFTRAM follows a spiral model: it iteratively collects scenarios, analyzes attribute interactions, identifies tradeoff and sensitivity points, and guides the refinement of framework structures and decisions.

\begin{figure}[ht]
    \centering
    \includegraphics[width=\textwidth]{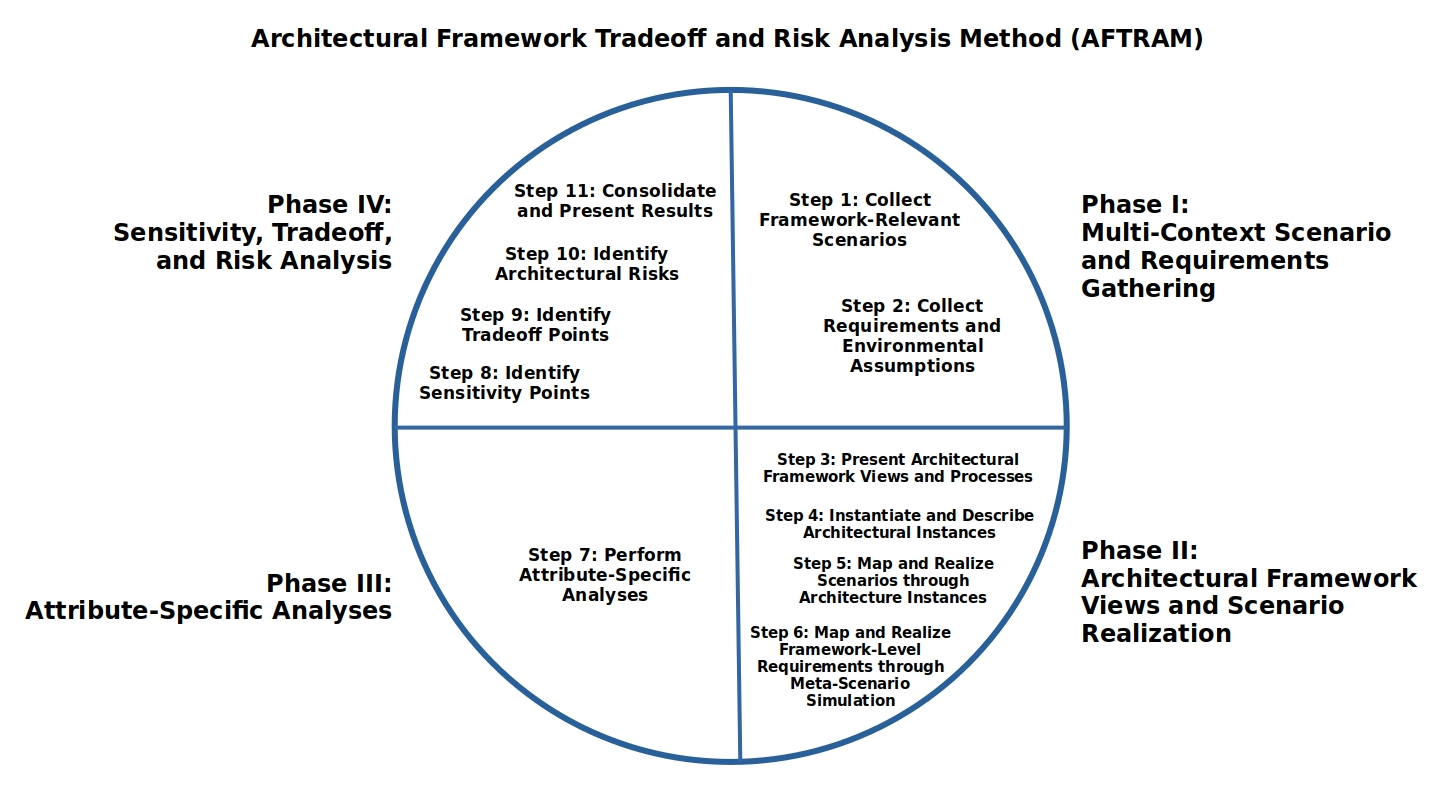}
    \caption{Steps of the Architectural Framework Tradeoff and Risk Analysis Method (AFTRAM)}
    \label{fig:aftram-spiral-model}
\end{figure}

\subsection{Evaluation Phases}

The AFTRAM method, similar to ATRAM and RATRAM, organizes its evaluation process into four phases, with refined artifacts and evaluation targets tailored to the context of architectural frameworks. Specifically, the names of Phase I and Phase II have been adjusted to emphasize alignment with framework-level considerations: \textbf{Phase I: Multi-Context Scenario and Requirements Gathering}, in which scenarios relevant to the framework and its potential derived systems are meticulously identified and documented; \textbf{Phase II: Architectural Framework Views and Scenario Realization}, which involves presenting the framework through multiple architectural viewpoints and mapping scenarios to assess its functional and structural adequacy; \textbf{Phase III: Attribute-Specific Analyses}, focusing on the evaluation of quality attributes such as performance and security within the specific framework context; and \textbf{Phase IV: Sensitivity, Tradeoff, and Risk Analysis}, which synthesizes the findings to identify critical sensitivities, trade-offs, and risks, guiding necessary architectural refinements. Together, these four phases are structured into 11 iterative steps, as illustrated by the spiral model in Figure~\ref{fig:aftram-spiral-model}, emphasizing the dynamic and cyclical nature of the evaluation process.

\subsubsection{Phase I: Multi-Context Scenario and Requirements Gathering}

\textbf{Purpose:}
To initiate the evaluation of an \textbf{Architectural Framework (AF)} by collecting scenarios and requirements across multiple system types, domains, and development processes. AFs define both \textbf{structural models and development methodologies}, so this phase includes structural, procedural, and lifecycle-focused evaluation. Scenario elicitation and requirements gathering are intertwined and iterative.

\textbf{Step 1: Collect Framework-Relevant Scenarios}
\begin{itemize}
    \item \textbf{Elicit scenarios addressing framework use and evolution}:
    \begin{itemize}
        \item Functional scenarios (core services and capabilities provided by the framework)
        \item Evolution scenarios (adapting framework to new needs/domains)
        \item Stress scenarios (boundary/exceptional usage)
        \item Adoption scenarios (initial and ongoing framework adoption by teams/orgs)
        \item Extension scenarios (how easily and cleanly the framework can be extended)
        \item Interoperability Scenarios
        \item Process scenarios (support for agile, model-driven, iterative workflows)
        \item Reference Architecture Creation Scenarios (how the framework enables new RA development)
    \end{itemize}
\end{itemize}

\textbf{Stakeholders:}
\begin{itemize}
    \item Framework designers (designers)
    \item System and reference architects (implementers)
    \item Process engineers, domain modelers (domain experts)
    \item Lifecycle owners, DevOps leads (operational roles)
\end{itemize}

\textbf{Step 2: Collect Requirements and Environmental Assumptions}
\begin{itemize}
    \item \textbf{Framework-Level Requirements:}
    \begin{itemize}
        \item Structural extensibility, methodology support, configurability, lifecycle alignment
    \end{itemize}

    \item \textbf{Derived-System Requirements:}
    \begin{itemize}
        \item Functional and quality requirements enabled in architectures derived from the framework
        \item Traceability to domain-wide goals
    \end{itemize}

    \item \textbf{Environmental Assumptions:}
    \begin{itemize}
        \item Regulatory and compliance conditions, toolchain expectations, platform dependencies, governance and lifecycle models
    \end{itemize}
\end{itemize}

\textbf{Phase I Artifacts:}

\begin{tabular}{|p{4cm}|p{6cm}|p{5cm}|}
\hline
Artifact & Description & Creation Process \\
\hline
Stakeholder \& Domain Map & Captures framework stakeholders and organizational contexts. & Interviews, role mapping \\ \hline
Scenario Catalog & Functional, evolution, stress, adoption, extension, process, and RA creation scenarios. & Multi-role workshops \\ \hline
Prioritized Scenario List & Focused scenario set for deeper exploration. & Risk-based selection \\ \hline
Framework-Level Requirement Document & Documents structure, methodology support, lifecycle integration. & Joint creation from scenarios \\ \hline
Derived-System Requirement Document & System-level expectations for architectures built using the framework. & Traced from usage scenarios \\ \hline
Environmental Assumptions Document & Records context-specific assumptions and constraints. & Documented collaboratively \\ \hline
Utility Tree & Maps quality concerns to framework scenarios. & Built interactively \\
\hline
\end{tabular}

\subsubsection{Phase II: Architectural Framework Views and Scenario Realization}

\textbf{Purpose:}
To describe the architectural framework’s structures, variability mechanisms, and lifecycle processes to establish a concrete basis for evaluation. It then proceeds to instantiate one or more candidate software architectures, motivated by anticipated scenario coverage and quality attribute support but without yet performing detailed realization. These instantiated architectures are subsequently mapped against the elicited scenarios to analyze realization strategies, classify support levels, and assess framework flexibility.

\textbf{Step 3: Present Architectural Framework Views and Processes}

\begin{itemize}
    \item \textbf{Description:}
    \begin{itemize}
        \item Describe the architectural framework using multiple viewpoints that reflect both its structural and methodological roles:
        \begin{itemize}
            \item \textbf{Structural Viewpoint}: framework-defined components, module types, and architectural structures
            \item \textbf{Interaction Viewpoint}: coordination mechanisms and data/control flow models between extensible architectural elements
            \item \textbf{Behavioral Viewpoint}: expected execution-time behaviors or runtime protocols that shape framework-conformant systems
            \item \textbf{Variability Viewpoint}: framework-supported extension mechanisms, configuration models, and areas of intended customization
            \item \textbf{Process Viewpoint}: methodology and lifecycle guidance provided by the framework, including process alignment (e.g., agile, iterative, model-driven)
        \end{itemize}
        \item Document the architectural constraints, extension points, and usage guidance embedded in the framework.
        \item Describe the rationale for major decisions in the framework’s structure and process support, particularly as they relate to enabling adaptability, reusability, and consistency across diverse system families.
    \end{itemize}
\end{itemize}

\textbf{Artifacts:}

\begin{tabular}{|p{4cm}|p{6cm}|p{5cm}|}
\hline
Artifact & Description & Creation Process \\
\hline
Scenario Realization Document & Documents how scenarios are realized through architectural instances, including realization strategies, support classifications, and analysis of any gaps or limitations introduced by the architectural framework. & Created through scenario-to-architecture mapping sessions and instance walkthroughs. \\
\hline
\end{tabular}

\textbf{Step 4: Instantiate and Describe Architectural Instances}

\begin{itemize}
    \item \textbf{Instantiate one or more candidate software architectures} based on the architectural framework’s structural elements, interaction models, extension mechanisms, and lifecycle processes. These instances illustrate how the framework supports the development of diverse systems or domains.

    \item \textbf{For each instantiated architecture:}
    \begin{itemize}
        \item \textbf{Describe the structure:}
        \begin{itemize}
            \item Identify the core components and configuration decisions made during instantiation
            \item Indicate which elements are:
            \begin{itemize}
                \item \textbf{Reused directly} (from the architectural framework with no modification)
                \item \textbf{Reused via supported extension mechanisms} (customized through framework-defined variation or extension points)
                \item \textbf{Extended or introduced} (beyond the framework’s intended scope or patterns)
            \end{itemize}
        \end{itemize}
        \item \textbf{Capture instantiation details:}
        \begin{itemize}
            \item Summarize the extension points exercised and how they were used
            \item Document any assumptions, constraints, or deviations made in aligning the architecture with framework guidance

        \end{itemize}
        \item \textbf{Map instantiation to framework structures and processes:}
        \begin{itemize}
            \item Clarify how structural decisions correspond to framework-provided elements such as architectural building blocks, templates, or development workflows
            \item Highlight any process alignment or methodology integration (e.g., model-driven development, agile practices, iterative refinement)

        \end{itemize}
        \item \textbf{Motivate the instantiated architecture:}
        \begin{itemize}
            \item Identify the scenarios this architecture intends to address and justify its alignment with framework-level quality attribute goals
            \item Provide rationale for architectural choices made in response to those scenarios
            \item Discuss tradeoffs, adaptations, or constraints that influenced the instantiation process
        \end{itemize}
    \end{itemize}
\end{itemize}

\textbf{Artifacts:}

\begin{tabular}{|p{4cm}|p{6cm}|p{5cm}|}
\hline
Artifact & Description & Creation Process \\
\hline
Architectural Instances Catalog & Document instantiated architectures derived from the framework, including their structure, configuration decisions, motivations for scenario support, and extension points used. & Developed collaboratively with architecture teams applying the framework in context. \\
\hline
\end{tabular}

\textbf{Step 5: Map and Realize Scenarios through Architecture Instances}

\begin{itemize}
    \item \textbf{For each scenario} from the Scenario Catalog (Phase I), identify one or more architecture instances that are capable of realizing the scenario.

    \item \textbf{Map each scenario to architectural instances} by:
    \begin{itemize}
        \item Tracing how framework-provided structures and extension mechanisms enable the realization within each instance
        \item Identifying the architectural elements involved in supporting the scenario
        \item Establishing traceability from framework-level structures through the instantiated architecture to the scenario flow

    \end{itemize}
    \item \textbf{Describe the realization strategy:}
    \begin{itemize}
        \item Detail how the scenario is handled within the instantiated architecture, including structural configurations and interaction models
        \item Highlight the extension points exercised, configuration paths followed, and quality attributes addressed

        \item Capture assumptions about behavior, environment, and usage conditions
    \end{itemize}
    \item \textbf{Classify the scenario support type:}
    \begin{itemize}
        \item \textbf{Natively Supported} --- scenario is fully realizable with minimal effort using default framework constructs
        \item \textbf{Guided Realization} --- scenario is supported via recommended framework extensions or documented practices
        \item \textbf{Constrained Realization} --- realization is feasible but requires workaround, deviation, or significant effort
        \item \textbf{Unsupported} --- framework limitations prevent realistic realization of the scenario

    \end{itemize}
    \item \textbf{Identify and analyze realization gaps:}
    \begin{itemize}
        \item Note any framework-level structural, behavioral, or process constraints that hinder scenario realization
        \item Document gaps in flexibility, extensibility, or configurability
        \item Recommend framework adaptations or enhancements where scenario needs are not sufficiently met
    \end{itemize}
\end{itemize}

\textbf{Artifacts:}

\begin{tabular}{|p{4cm}|p{6cm}|p{5cm}|}
\hline
Artifact & Description & Creation Process \\
\hline
Scenario Realization Document & Map scenarios to architectural instances, describe realization strategies, classify realization types, and embed traceability from framework elements through architecture to scenario behavior. & Produced via structured walkthroughs of instantiated architectures and framework mappings. \\
\hline
\end{tabular}

\textbf{Step 6: Map and Realize Framework-Level Requirements through Meta-Scenario Simulation}

- This step evaluates how the architectural framework supports \textbf{framework-level requirements} by simulating abstract, forward-looking usage situations called \textbf{meta-scenarios}.

- \textbf{Meta-scenarios} differ from the system-level scenarios analyzed in Step 3. Rather than describing specific application behavior, they explore how the framework itself responds to broader concerns such as reuse across unforeseen domains, evolution under changing constraints, or integration with alternative processes and toolchains.

- \textbf{For each framework-level requirement}, analyze how the framework structure, extension mechanisms, and process guidance support, partially support, or fail to support the realization of relevant meta-scenarios.

- \textbf{Simulate meta-scenarios} that illustrate how the framework addresses challenges related to generality, flexibility, and long-term sustainability. For each case, describe how the framework enables, constrains, or conditions the realization of the requirement under evaluation.

- \textbf{Map requirements to framework structures and processes} by:
    \begin{itemize}
        \item Identifying the architectural and process-level elements intended to support or influence each framework-level requirement as it applies in a meta-scenario context
        \item Tracing the relevant mechanisms such as extension points, configuration strategies, and lifecycle integration models that contribute to requirement satisfaction
    \end{itemize}

- \textbf{Classify realization support level for each framework-level requirement:}
    \begin{itemize}
        \item \textbf{Fully Met} --- requirement is fully addressed through framework constructs or process mechanisms
        \item \textbf{Partially Met} --- requirement is conditionally supported, requiring additional effort or assumptions
        \item \textbf{Not Met} --- requirement cannot be realistically supported under current framework constraints
    \end{itemize}

- \textbf{Identify realization gaps and constraint boundaries:}
    \begin{itemize}
        \item Analyze structural or methodological limitations that reduce framework flexibility or generalizability
        \item Recommend potential extensions, generalizations, or tool/process integrations that could improve framework-level support
    \end{itemize}

\textbf{Artifacts:}

\begin{tabular}{|p{4cm}|p{6cm}|p{5cm}|}
\hline
Artifact & Description & Creation Process \\
\hline
Meta-Scenario Realization Document & Documents the realization of framework-level requirements through meta-scenario simulations, maps supporting framework structures and processes, classifies requirement satisfaction levels, and identifies gaps or limitations. & Simulated by analysis teams using abstract test cases and meta-scenario walkthroughs. \\
\hline
\end{tabular}

\subsubsection{Phase III: Attribute-Specific Analyses}

\textbf{Purpose:}
To analyze each quality attribute in isolation with respect to the architectural framework’s structures, extension mechanisms, and process guidance. This phase supports reasoning about attribute satisfaction \textbf{across diverse system families and usage contexts}, without yet addressing tradeoffs or inter-attribute interactions.

\textbf{Step 7: Perform Attribute-Specific Analyses}

Each relevant quality attribute is analyzed separately using framework viewpoints, scenario realizations, and instantiated architectures from Phase II. Evaluation methods may involve reasoning about process-structure interactions, generality constraints, and extensibility mechanisms. The goal is to determine how well the framework enables or limits attribute realization under its supported development approaches. All evaluations are conducted independently to preserve separation of concerns.

\textbf{Artifacts:}

\begin{tabular}{|p{4cm}|p{6cm}|p{5cm}|}
\hline
Artifact & Description & Creation Process \\
\hline
Attribute Evaluation Report & Attribute-specific assessment of how the architectural framework supports each quality attribute across contexts. Includes evaluation rationale, assumptions, structure/process considerations, and generalization notes. & Developed using attribute walkthroughs, instantiated architecture evidence, and framework guideline review. \\
\hline
\end{tabular}

\subsubsection{Phase IV: Sensitivity, Tradeoff, and Risk Analysis}

AFTRAM Phase IV is structurally identical to ATRAM Phase IV, reusing the same analytical flow while adapting it to the broader scope of architectural frameworks. Sensitivity and tradeoff points often occur at the intersection of structural and methodological elements---such as extensibility mechanisms, lifecycle models, or configuration strategies---while risks may involve limited generalizability, rigid process coupling, or assumptions that hinder adaptation across domains. The evaluation focuses on how well the framework balances consistency, flexibility, and attribute support across its intended range of use cases.

\subsubsection{Iteration and Refinement in AFTRAM}

Like ATAM and ATRAM, AFTRAM follows a spiral model, enabling iterative refinement across phases. As attribute evaluations or tradeoff points reveal framework limitations or misalignments, adjustments to structure, extension mechanisms, or process guidance may prompt re-entry into earlier phases for continued analysis.

\subsection{Case Study: Remote Monitoring Architectural Framework (RMAF)}

To validate the applicability of AFTRAM, the Remote Monitoring Architectural Framework (RMAF) was specifically crafted as an example to evaluate AFTRAM's application to architectural frameworks. RMAF is a flexible and comprehensive methodology aimed at supporting the development of remote monitoring systems across diverse domains, including industrial monitoring, healthcare, and environmental sensing. The framework defines reusable building blocks, interaction models, and extensibility mechanisms for creating both domain-specific reference architectures and concrete system architectures. (For a more comprehensive description of the RMAF, see Appendix~\ref{app:rmaf-description}).

\noindent \textbf{Note:} This is an ongoing work. Future versions of this paper will incorporate extensive case-specific evaluations of the RMAF using the Architectural Framework Tradeoff Analysis Method (AFTRAM).


\section{Cross-Method Synthesis and Lessons Learned}
\label{sec:cross-method-synthesis-and-lessons-learned}

This section systematically compares all the methods developed in the context of ATRAF: ATRAM, RATRAM, and AFTRAM, through various aspects. The goal of this comparison is to highlight the distinct challenges and solutions associated with each method, as well as how they contribute to a comprehensive evaluation framework across different levels of architectural abstraction. By examining the methods in terms of their target scope, stakeholder involvement, viewpoints used, scenario realization, tradeoff analysis, and risk identification, we aim to provide a deeper understanding of how ATRAF supports the evaluation of software architectures, reference architectures, and architectural frameworks. This comparison serves to clarify the unique strengths and applications of each method, offering valuable insights for practitioners navigating the complexities of architectural decision-making.

\subsection{Comparative Summary of ATRAM, RATRAM, and AFTRAM}

\begin{table}[ht]
    \centering
    \caption{Comparative Table: ATRAM vs. RATRAM vs. AFTRAM}
    \label{tab:atram_ratram_aftram_comparison}
    \begin{tabular}{p{1.5cm}|p{0.28\textwidth}|p{0.31\textwidth}|p{0.31\textwidth}}
        \toprule
        \textbf{Aspect} & \textbf{ATRAM} (Architecture) & \textbf{RATRAM} (Reference Architecture) & \textbf{AFTRAM} (Architectural Framework) \\
        \midrule
        \textbf{Target} & Concrete \textbf{software architecture} & Generalized \textbf{reference architecture} & Meta-level \textbf{architectural framework} \\ \hline
        \textbf{Abstraction Level} & Low (system-specific) & Medium (domain-specific, reusable) & High (cross-domain, methodology-oriented) \\ \hline
        \textbf{Scenario Scope} & Functional, evolution, and stress scenarios at the \textbf{system level} & Adds \textbf{adoption} and \textbf{interoperability} scenarios for domain reuse & Adds \textbf{extension}, \textbf{process}, and \textbf{meta-scenarios} for framework lifecycle and generality \\ \hline
        \textbf{Stakeholders} & Architects, developers, users & RA designers, system architects, domain experts & Framework designers, reference/system architects, process engineers \\ \hline
        \textbf{Viewpoints Used} & Structural, interaction, behavioral, deployment & Adds \textbf{variability viewpoint} & Adds \textbf{process viewpoint}, integration/extension guidance \\ \hline
        \textbf{Scenario Realization} & Maps concrete scenarios to architecture components and paths & Maps \textbf{generalized scenarios} via instantiations derived from RA & Maps \textbf{framework-level and derived-level scenarios} via architecture families and meta-scenarios \\ \hline
        \textbf{Tradeoff Analysis Level} & Between quality attributes in a specific system & At domain level across multiple instantiations & Across process, structure, and extension mechanisms in diverse system families \\ \hline
        \textbf{Risk Identification} & System-specific risks (e.g., latency, fragility) & Risks in reuse, misalignment of domain patterns & Risks in process rigidity, extensibility constraints, lack of generality \\ \hline
        \textbf{Instance Role} & Evaluation focuses on the \textbf{architecture itself} & Evaluation involves creating/assessing \textbf{system instances from the RA} & Evaluation involves \textbf{system and RA instances generated from the framework} \\ \hline
        \textbf{Evaluation Method Adaptation} & Extends ATAM (1998 \& 2000) with explicit risk phase and iterative spiral process & Adapts ATRAM + incorporates context multiplicity, abstract property focus (from ATAM/R) & Adapts RATRAM + evaluates process support, methodology flexibility, and generalizability \\ \hline
        \textbf{Output Artifacts} & Utility tree, sensitivity/tradeoff/risk lists, evaluation summary & Adds \textbf{instantiated architectures}, scenario realization mappings, and reuse classifications & Adds \textbf{meta-scenario realizations}, process traceability, and generalization constraint analyses \\ \hline
        \textbf{Evaluation Focus} & Achieving system-level quality goals & Supporting domain-wide consistency, reuse, and adaptability & Enabling system family creation, framework evolution, and balance between flexibility and structure \\ \hline
        \textbf{Example from RTS Case} & RTSA (Remote Temperature Sensor Architecture) & RTSRA (Remote Temperature System Reference Architecture) & RMAF (Remote Monitoring Architectural Framework) \\
        \bottomrule
    \end{tabular}
\end{table}

\subsection{Insights from the RTS-Based Example Chain}

As we continue to evaluate ATRAM, RATRAM, and AFTRAM through the Remote Temperature Sensor (RTS) case family, future versions of this paper will provide a detailed exploration of the specific lessons learned from applying these methods to the case study. This ongoing work will address how each method contributes to the iterative refinement of architectural artifacts at different abstraction levels and will offer further insights into the practical applications and challenges of using ATRAF in real-world scenarios. Stay tuned for more in-depth analyses in future versions of the preprint.


\section{Related Work}
\label{sec:related-work}

The evaluation of software architectures has been a key research focus for several decades, leading to the development of a variety of methods that aim to assess how well an architecture supports desired quality attributes and business goals. Early contributions in this domain, such as the \textbf{Software Architecture Analysis Method (SAAM)}~\cite{kazman1994saam}, introduced the concept of scenario-based evaluation for software architectures. SAAM pioneered the use of \textit{quality-focused scenarios}, which were utilized to assess an architecture's ability to meet specific requirements, particularly in terms of modifiability. While effective in identifying the potential weaknesses of an architecture when it comes to a single quality attribute, SAAM was limited in its ability to handle tradeoffs between multiple quality attributes within a given architecture.

The \textbf{Architecture Tradeoff Analysis Method (ATAM)}~\cite{kazman2000atam}, represented a significant extension of SAAM by introducing a structured, \textit{stakeholder-centric} approach for evaluating multiple quality attributes simultaneously. ATAM employs a \textit{utility tree} to categorize and prioritize quality concerns and uses \textit{scenario-based evaluation} to assess how architectural decisions affect each of these concerns. It helps architects and stakeholders understand the tradeoffs between conflicting quality attributes, such as performance, security, and modifiability. Key outcomes of ATAM include the identification of \textit{sensitivity points}---decisions that have a significant impact on a particular quality---and \textit{tradeoff points}, where enhancing one attribute may negatively affect another. ATAM has become one of the most widely used and influential methods for software architecture evaluation, thanks to its repeatable process and comprehensive handling of multiple quality attributes \cite{clements2002evaluatingsa}.

Several adaptations and extensions of ATAM have been proposed over the years to address its limitations and to expand its applicability to new contexts. One of these is the \textbf{Cost Benefit Analysis Method (CBAM)}, which was introduced to incorporate \textit{economic reasoning} into architectural decision-making. CBAM complements ATAM by allowing decision-makers to assess the \textit{economic implications} of architectural choices, integrating cost, benefit, and return-on-investment calculations to prioritize trade-offs in business terms~\cite{kazman2001cbam}. This extension was particularly useful for decision-makers who needed to balance technical concerns with economic considerations.

Other notable extensions of ATAM focus on \textit{product-line architectures}, where an architecture must be able to accommodate a family of related products rather than a single system. \textbf{Extended ATAM (EATAM)} was proposed by \cite{kim2008eatam} to evaluate software product lines, incorporating variability analysis and \textit{variation points} that are crucial in product-line architectures. This extension retains the core principles of ATAM but adds new steps to analyze \textit{variability scenarios}, making it possible to evaluate how an architecture can support different configurations within a product family. Similarly, the \textbf{Holistic Product Line Architecture Assessment (HoPLAA)}~\cite{olumofin2007holistic} takes a two-phase approach, first evaluating the \textit{reference architecture} and then analyzing how this architecture impacts specific products in the line. These product-line-specific extensions demonstrate how the basic principles of ATAM can be adapted to handle reuse and variability, important concerns in modern software engineering.

While ATAM and its extensions have been broadly applied to concrete software architectures and product-line architectures, challenges persist in evaluating \textit{reference architectures} and \textit{architectural frameworks}, which are inherently more abstract and general. Reference architectures, which serve as reusable templates for domain-specific systems, must address a wider range of stakeholders and quality attributes across multiple system instantiations. However, ATAM was designed primarily for concrete, system-specific evaluations and is not directly applicable to these more abstract forms.

Angelov et al. \cite{angelov2014atamr} \cite{angelov2014atamr} addressed this gap by adapting ATAM for \textit{reference architecture evaluation}. Their work recognizes that reference architectures, unlike concrete architectures, are intended for reuse across multiple systems and therefore require a method that can handle \textit{context multiplicity}---the ability to assess an architecture across different deployment scenarios and use cases. They introduced an adapted version of ATAM that adjusts the stakeholder selection process, scenario elicitation, and utility prioritization to better accommodate the broad scope and \textit{reusability} inherent in reference architectures. This adaptation emphasizes the need for specialized evaluation approaches when dealing with \textit{architectures of many systems}.

Further complicating the evaluation process, \textit{architectural frameworks}---which provide methodologies, principles, and patterns for designing families of systems---present unique challenges that ATAM and its variants are not equipped to address. Architectural frameworks are typically highly abstract and flexible, designed to support the development of multiple system families across different domains. Evaluating such frameworks requires an approach that goes beyond single-system or domain-specific analysis to account for the \textit{lifecycle evolution} and \textit{long-term adaptability} of the systems they guide.

In response to these challenges, the proposed \textbf{Architectural Framework Tradeoff and Risk Analysis Method (AFTAM)} aims to extend ATAM's scenario-based evaluation framework to address the specific concerns of \textit{architectural frameworks}. By focusing on \textit{meta-scenarios} that address the lifecycle and long-term adaptability of frameworks, AFTAM seeks to evaluate the \textit{flexibility} and \textit{scalability} of frameworks while maintaining consistency across diverse system families. This approach builds on the foundational work of ATAM and its extensions, but tailors it to the unique needs of frameworks and the broader scope of evaluation they demand.

In summary, while numerous methods have been developed for evaluating concrete software architectures, reference architectures, and product-line systems, there remains a significant gap in the evaluation of \textit{architectural frameworks}. This research aims to fill this gap by proposing a comprehensive approach that integrates the strengths of ATAM with the unique challenges presented by frameworks, offering a systematic way to evaluate trade-offs and risks across different levels of architectural abstraction.


\section{Conclusion and Future Work}
\label{sec:conclusion}

In this paper, we introduced the Architecture Tradeoff and Risk Analysis Framework (ATRAF), a unified approach for evaluating software architectures, reference architectures, and architectural frameworks. This work expands upon existing methods like the Architecture Tradeoff Analysis Method (ATAM) and its extensions, such as ATAM/R, by providing tailored evaluation techniques for each level of abstraction. The paper's primary contributions include the design and application of ATRAF's three evaluation methods---ATRAM, RATRAM, and AFTRAM---each addressing the unique challenges of system-specific, domain-specific, and framework-level architectures, respectively. Through a hierarchical structuring of architectural concepts, we highlighted the distinctions between software architectures, reference architectures, and architectural frameworks, showing how each requires different evaluation strategies to uncover tradeoffs and risks effectively.

Looking ahead, future iterations of this work will delve into detailed case studies across a diverse range of domains, providing deeper insights into how ATRAF can be applied to guide decision-making in real-world scenarios. These case studies will enhance our understanding of ATRAF's utility across a wide spectrum of architectural artifacts, from concrete software architectures to highly flexible architectural frameworks. By analyzing how real-world systems instantiate frameworks, we aim to uncover the nuances involved in managing architectural tradeoffs, ultimately strengthening ATRAF’s applicability and robustness.

Although ATRAF shows great promise, it remains a work in progress. In fact, several obstacles still need to be addressed, particularly around ensuring scenario completeness, managing stakeholder diversity, and overcoming the abstraction barriers inherent in frameworks. Additionally, the complexity of attribute interdependencies and the resource-intensive nature of comprehensive evaluations pose further challenges. Future work will focus on refining the evaluation process, expanding the set of quality attributes to incorporate emerging concerns such as sustainability and long-term adaptability. Furthermore, significant efforts will be directed toward improving tool support for scenario management and tradeoff visualization, as well as incorporating quantitative models for better prioritization and communication of tradeoffs.

This paper lays the groundwork for the continued development and refinement of ATRAF, aiming to provide the software architecture community with a powerful tool for evaluating and managing the complexities of tradeoffs and risks in modern software systems. Future research will focus on further validating and extending ATRAF’s methods, ensuring that the framework evolves in parallel with the growing complexity of software systems and their architectures.


\appendix

\section{Appendix: Examples of Architectural Artifacts}
\label{app:example-of-architectural-artifacts}

This appendix provides detailed architectural descriptions of the three example systems used throughout the paper: RTSA, RTSRA, and RMAF. These artifacts demonstrate how the Architecture Tradeoff and Risk Analysis Framework (ATRAF) applies across different levels of architectural abstraction.

\subsection{Remote Temperature Sensor Architecture (RTSA)}
\label{app:rtsa-description}

The \textbf{Remote Temperature Sensor Architecture (RTSA)} specifies the design of a system that remotely monitors furnace temperatures, fulfilling both functional and quality requirements such as performance, availability, and security.

RTSA defines around three main entities:
\begin{itemize}
    \item \textbf{Furnaces}: Physical units providing temperature data.
    \item \textbf{Server}: Central system managing data acquisition and communication.
    \item \textbf{Clients}: Operator systems receiving updates and sending control requests.
\end{itemize}

The \textbf{Server} contains:
\begin{itemize}
    \item \textbf{Analog-to-Digital Converter (ADC) Module}: Converts furnace temperature readings to digital values, handling one reading at a time.
    \item \textbf{Communication Interface Module}: Manages client-server messaging, including sending periodic updates and processing control requests.
    \item \textbf{Scheduling Tasks}: One per furnace, each responsible for periodically reading a furnace’s temperature, processing it through the ADC, and sending updates to the corresponding client. Each task's frequency can be dynamically adjusted.
\end{itemize}

The system operates by allowing clients to configure update rates individually for each furnace. The server schedules temperature acquisitions accordingly and forwards updates, ensuring bounded latencies, predictable jitter, and resilience against failures.

To ensure secure communication, a secure channel must be established between the server and clients. Specifically, HTTPS must be used when communication is based on HTTP, and secure Secure Web Socket (WSS) must be used when WebSocket connections are employed.

RTSA emphasizes early handling of quality tradeoffs, considering performance under load, system availability during failures, and basic security against message tampering.

\subsection{Remote Temperature System Reference Architecture (RTSRA)}
\label{app:rtsra-description}

The \textbf{Remote Temperature System Reference Architecture (RTSRA)} provides a reusable architectural template for designing systems that perform remote temperature monitoring across various domains, including but not limited to furnace monitoring, air conditioning systems, and environmental sensing. RTSRA supports the systematic creation of systems like RTSA by defining a structured set of entities, components, modules, and interaction patterns aligned with explicit quality requirements.

RTSRA defines three core architectural entities:
\begin{itemize}
    \item \textbf{Temperature Sources}: Abstract sensors providing temperature data. These sources are not tied to any specific type of equipment (e.g., furnaces) and may represent any monitored asset.
    \item \textbf{Server(s)}: Processing units responsible for acquiring sensor data, managing control commands, and communicating with clients. The architecture allows for one or multiple servers per deployment, enabling redundancy and fault tolerance.
    \item \textbf{Clients}: Operator systems or automated applications that issue control requests and consume temperature updates.
\end{itemize}

\subsubsection{Server Component Structure}

The \textbf{Server} in RTSRA includes the following modules:
\begin{itemize}
    \item \textbf{ADC Interface Module}: Abstracts the acquisition of temperature readings from sensors, handling single access at a time.
    \item \textbf{Communication Interface Module}: Manages bidirectional client communication, supporting only \textbf{HTTP} and \textbf{WebSocket} protocols. Protocols requiring intermediaries, such as \textbf{MQTT}, are outside the scope of RTSRA.
    \item \textbf{Scheduling Task Modules}: One task per temperature source, each responsible for periodically initiating a reading, processing the result through the ADC, and sending updates to the assigned client. Task execution frequencies are configurable by clients.
\end{itemize}

\subsubsection{Redundancy and Fault Tolerance}

RTSRA accommodates both non-redundant and redundant server configurations. In redundant setups, it provides a \textbf{default failover algorithm} to ensure service continuity:
\begin{itemize}
    \item Clients monitor server responsiveness.
    \item Upon detecting server failure (e.g., missed heartbeats, communication timeouts), a client automatically switches to a designated backup server.
    \item The default failover behavior is specified but must be manually customized if designers require non-standard recovery policies (e.g., load balancing, prioritization).
\end{itemize}

Customization of the fault tolerance mechanism is \textbf{manual}, requiring changes at the design or code level.

\subsubsection{Interaction Model}

A typical interaction in an RTSRA-based system follows these steps:
\begin{enumerate}
    \item A \textbf{client} sends a \textbf{control request} to a \textbf{primary server}, specifying the desired frequency for updates from a particular sensor.
    \item The \textbf{server} adjusts the corresponding \textbf{scheduling task} to reflect the new frequency.
    \item The \textbf{scheduling task} periodically triggers the \textbf{ADC interface} to acquire temperature readings and forwards the processed data to the client via the \textbf{communication interface}.
    \item \textbf{Fault Handling}:
    \begin{itemize}
        \item Clients continuously monitor the availability of their connected server.
        \item If the \textbf{primary server} becomes unresponsive (due to hardware or software faults), the client automatically switches to a \textbf{backup server}, based on the default failover algorithm.
        \item The backup server resumes responsibility for handling control requests and delivering updates, minimizing disruption and maintaining system availability.
    \end{itemize}
\end{enumerate}

This model ensures predictable behavior during normal operation and controlled recovery during failure scenarios.

RTSRA balances structure and flexibility, promoting the reuse of proven architectural patterns while enabling adaptation to different operational contexts, all while maintaining quality attribute goals such as \textbf{availability}, \textbf{modifiability}, and \textbf{scalability}.

\subsection{Remote Monitoring Architectural Framework (RMAF)}
\label{app:rmaf-description}

The \textbf{Remote Monitoring Architectural Framework (RMAF)} is a comprehensive architectural framework designed to support the systematic development of remote monitoring systems across diverse domains and application contexts. It provides a \textbf{methodology}, \textbf{process structures}, \textbf{guidelines}, and \textbf{extensibility mechanisms} that enable architects to create both domain-specific reference architectures (such as RTSRA) and concrete system architectures (such as RTSA).

RMAF emphasizes flexibility, scalability, and adaptability, recognizing that remote monitoring needs vary widely---from industrial temperature sensing to environmental monitoring, healthcare telemetry, and system performance tracking. As a framework, RMAF does not prescribe a single fixed architecture but defines reusable building blocks, viewpoints, and design principles that guide consistent architecture development across families of remote monitoring systems.

\subsubsection{RMAF Structure and Components}

RMAF is organized around several key modeling structures:
\begin{itemize}
    \item \textbf{Monitoring Sources}: Abstract entities representing any measurable system, asset, or environment component (e.g., temperature sensors, vibration monitors, network performance counters, patient heart rate monitors). Monitoring sources may vary in capabilities, criticality, and connectivity.
    \item \textbf{Processing Nodes}: Components that ingest, process, and manage monitoring data. They may be centralized (single server model), decentralized (distributed nodes), or hierarchical (aggregated regional servers).
    \item \textbf{Client Applications}: Systems that consume monitoring data, issue control commands, visualize system status, and trigger automated responses based on configurable rules.
    \item \textbf{Communication Mechanisms}: Abstracted to support multiple protocols (e.g., HTTP, WebSocket, MQTT, AMQP) with selectable reliability and security guarantees based on system needs.
    \item \textbf{Scheduling and Control Models}: Framework guidelines for defining periodic, event-driven, or adaptive data acquisition models. RMAF provides patterns for dynamically adjusting sampling rates, prioritizing monitoring sources, and managing control feedback loops.
\end{itemize}

\subsubsection{Core Principles and Methodologies}

RMAF defines a consistent set of core methodologies:

\begin{itemize}
    \item \textbf{Separation of Concerns}: Clear separation between data acquisition, data processing, control management, and client interaction to promote modifiability and scalability.

    \item \textbf{Quality Attribute Focus}: Tradeoff models guiding design decisions to balance critical qualities such as performance (low latency, high throughput), availability (redundant failover support), modifiability (dynamic source integration), and security (data integrity, confidentiality).

    \item \textbf{Extensibility and Customization}: Standardized extension points allow designers to introduce new sensor types, data processing algorithms, communication protocols, and client roles without destabilizing the system structure.

    \item \textbf{Redundancy and Fault Tolerance Patterns}: Guidance on selecting and implementing redundancy models (e.g., active-passive failover, load balancing) and failover mechanisms, either through client-driven detection (as in RTSRA) or system-managed orchestration.
\end{itemize}

Importantly, \textbf{RMAF tolerates and anticipates the inclusion of additional architectural entities}, such as brokers, caches, or analytics services, which may be required when integrating certain technologies or protocols. However, RMAF defines \textbf{requirements, constraints, and integration principles} to ensure that such additions do not compromise the framework’s quality attribute goals. For example, a designer may choose to implement an MQTT-based communication model that introduces a broker into the system. In this case, RMAF provides guidance to ensure that the broker’s inclusion adheres to security constraints (e.g., authentication, message integrity), availability expectations (e.g., broker failover), and performance requirements (e.g., low-latency messaging). These integration principles help preserve system-wide consistency, safety, and evolvability, even as architectures extend beyond the baseline framework.

\subsubsection{Relationship to RTSRA and RTSA}

\begin{itemize}
    \item \textbf{RMAF to RTSRA}: RTSRA is an example of a \textbf{reference architecture} instantiated from RMAF. It narrows RMAF’s broad methodology to a specific domain: remote temperature monitoring systems. RTSRA applies RMAF’s principles, selecting particular interaction models (e.g., HTTP/WebSocket), fault tolerance strategies (client-driven failover), and system roles (temperature sources, servers, clients).
    \item \textbf{RMAF to RTSA}: RTSA is an example of a \textbf{concrete system architecture} that could be created by further specializing RTSRA. It realizes the abstract structures of RTSRA into a specific implementation for furnace temperature monitoring, with defined ADC modules, scheduling tasks, communication handlers, and fault handling strategies.
\end{itemize}

Thus, RMAF enables the creation of diverse remote monitoring systems, offering the flexibility to:
\begin{itemize}
    \item Directly design new systems by applying its principles and guidelines.
    \item Develop domain-specific reference architectures like RTSRA.
    \item Support system-specific designs like RTSA derived from a reference architecture or directly guided by the framework.
\end{itemize}

\subsubsection{Example Usage of RMAF}

A development team tasked with building a cloud-based environmental monitoring platform could start from RMAF:
\begin{itemize}
    \item Define new types of \textbf{Monitoring Sources} (e.g., air quality sensors, water pollution meters).
    \item Tailor the \textbf{Processing Nodes} to handle high-volume real-time data streams.
    \item Select communication protocols emphasizing scalability (e.g., MQTT over WebSocket).
    \item Introduce a \textbf{broker} to support MQTT messaging, ensuring that its integration follows RMAF’s specified security and availability constraints.
    \item Apply redundancy patterns for data aggregation services using active-active server clusters.
    \item Extend scheduling models to include adaptive sampling rates based on real-time analysis of environmental events.
\end{itemize}

All while systematically managing tradeoffs between data latency, system availability, modifiability, and energy efficiency---guided by the structure and flexibility that RMAF provides.

\bibliographystyle{unsrtnat}
\bibliography{references}

\end{document}